\documentclass[fleqn,usenatbib]{mnras}

\usepackage{newtxtext,newtxmath}

\usepackage[T1]{fontenc}

\DeclareRobustCommand{\VAN}[3]{#2}
\let\VANthebibliography\thebibliography
\def\thebibliography{\DeclareRobustCommand{\VAN}[3]{##3}\VANthebibliography}

\usepackage{graphicx}	
\usepackage{amsmath}	
\usepackage{comment}
\usepackage{color}
\usepackage{xcolor}
\usepackage{soul}
\usepackage[toc,page]{appendix}

\newcommand{\epta}[1]{{EPTA~D.R.~1.0}}
\newcommand{\ipta}[1]{{IPTA~D.R.~1.0}}

\newcommand{\rmpi}{\mathrm{\pi}}
\newcommand{\e}{\mathrm{e}}

\newcommand{\C}{\textsf{\textbf{C}}}
\newcommand{\Cinv}{\textsf{\textbf{C}}^{-1}}
\newcommand{\Cda}{\textsf{\textbf{C}}'}
\newcommand{\Cdainv}{\textsf{\textbf{C}}'^{-1}}

\newcommand{\Ctot}{\tilde{\textsf{\textbf{C}}}}

\newcommand{\Cw}{\textsf{\textbf{C}}_{\textsf{W}}}
\newcommand{\Cr}{\textsf{\textbf{C}}_{\textsf{R}}}
\newcommand{\Cdm}{\textsf{\textbf{C}}_{\textsf{DM}}}
\newcommand{\Ccu}{\textsf{\textbf{C}}_{\textsf{CURN}}}

\newcommand{\M}{\textsf{\textbf{M}}}
\newcommand{\Mtr}{\textsf{\textbf{M}}^{\textsf{T}}}

\newcommand{\T}{\textsf{\textbf{T}}}
\newcommand{\Ttr}{\textsf{\textbf{T}}^{\textsf{T}}}
\newcommand{\B}{\textsf{\textbf{B}}}
\newcommand{\G}{\textsf{\textbf{G}}}

\newcommand{\Gtr}{\textsf{\textbf{G}}^{\textsf{T}}}

\newcommand{\BF}{\mathcal{B}}

\newcommand{\F}{\textsf{\textbf{F}}}
\newcommand{\Ftr}{\textsf{\textbf{F}}^{\textsf{T}}}

\newcommand{\tr}{\textsf{T}}

\newcommand{\res}{\delta\boldsymbol{t}}

\newcommand{\ttwo}{{\tt TEMPO2}}
\newcommand{\tn}{{\tt TEMPONEST}}
\newcommand{\mnest}{{\tt MULTINEST}}
\newcommand{\pmnest}{{\tt PYMULTINEST}}
\newcommand{\python}{{\tt PYTHON}}
\newcommand{\pchord}{{\tt POLYCHORD}}
\newcommand{\pchordlite}{{\tt POLYCHORDLITE}}
\newcommand{\psrx}{{\tt PSRCHIVE}}
\newcommand{\ftwo}{{\tt FORTYTWO}}
\newcommand{\beph}{{\tt BAYESEPHEM}}
\newcommand{\lmoss}{{\tt LINIMOSS}}
\newcommand{\egp}{{\tt EPHEMGP}}
\newcommand{\eprise}{{\tt ENTERPRISE}}
\newcommand{\ee}{{\tt ENTERPRISE\_EXTENSIONS}}
\newcommand{\ptmc}{{\tt PTMCMCSAMPLER}}

\definecolor{mygray}{gray}{0.6}

\title[EPTA common-red-signal analysis]{Common-red-signal analysis with 24-yr high-precision timing of the European Pulsar Timing Array: Inferences in the stochastic gravitational-wave background search}

\author[S. Chen et al.]{\parbox{\textwidth}{
S.~Chen$^{1,2}$\thanks{E-mail:siyuan.chen@cnrs-orleans.fr},
R.~N.~Caballero$^{3}$\thanks{E-mail:caballero.astro@gmail.com},
Y.~J.~Guo$^{4}$,
A.~Chalumeau$^{5,1,2}$,
K.~Liu$^{4}$,
G.~Shaifullah$^{6,7}$,
K.~J.~Lee$^{3,8,4}$,
S.~Babak$^{5,9}$,
G.~Desvignes$^{4,10}$,
A.~Parthasarathy$^{4}$,
H.~Hu$^{4}$,
E.~van~der~Wateren$^{11,12}$, 
J.~Antoniadis$^{13,4,14}$,
\newline
A.-S.~Bak~Nielsen$^{4,15}$,
C.~G.~Bassa$^{11}$,
A.~Berthereau$^{1,2}$,
M.~Burgay$^{16}$,
D.~J.~Champion$^{4}$,
I.~Cognard$^{1,2}$,
M.~Falxa$^{5}$,
R.~D.~Ferdman$^{17}$,
P.~C.~C.~Freire$^{4}$,
J.~R.~Gair$^{18}$,
E.~Graikou$^{4}$,
L.~Guillemot$^{1,2}$,
J.~Jang$^{4}$,
G.~H.~Janssen$^{11,12}$,
R.~Karuppusamy$^{4}$,
M.~J.~Keith$^{19}$,
M.~Kramer$^{4,19}$,
X.~J.~Liu$^{20,19}$,
A.~G.~Lyne$^{19}$,
R.~A.~Main$^{4}$,
J.~W.~McKee$^{21}$,
M.~B.~Mickaliger$^{19}$,
B.~B.~P.~Perera$^{22}$,
D.~Perrodin$^{16}$,
A.~Petiteau$^{5}$,
N.~K.~Porayko$^{4}$,
A.~Possenti$^{16,23}$,
A.~Samajdar$^{6}$,
S.~A.~Sanidas$^{19}$,
A.~Sesana$^{6,7}$,
L.~Speri$^{18}$,
B.~W.~Stappers$^{19}$,
G.~Theureau$^{1,2,24}$,
C.~Tiburzi$^{11}$,
A.~Vecchio$^{25}$,
J.~P.~W.~Verbiest$^{15,4}$,
J.~Wang$^{15}$,
L.~Wang$^{8}$
and H.~Xu$^{3,26,8}$}
\vspace{0.4cm} \\ 
Affiliations are at the end of the paper
}

\date{Accepted XXX. Received YYY; in original form ZZZ}

\pubyear{2021}

\begin{document}
\label{firstpage}
\pagerange{\pageref{firstpage}--\pageref{lastpage}}
\maketitle

\begin{abstract}

We present results from the search for a 
stochastic gravitational-wave background (GWB) as predicted by the theory of General Relativity using six radio millisecond pulsars
from the Data Release 2 (DR2) of the European Pulsar Timing Array (EPTA) covering a timespan up to 24 years.
A GWB manifests itself as a long-term low-frequency stochastic signal common to all pulsars, a common red signal (CRS), with the characteristic Hellings-Downs (HD) spatial correlation.
Our analysis is performed with two independent pipelines, \eprise{} and \tn{}+\ftwo{}, 
which produce consistent results.
A search for a CRS with
simultaneous estimation of its spatial correlations 
yields spectral properties compatible with theoretical GWB predictions,
but does not result in the required measurement of the HD correlation, as required for GWB detection.
Further Bayesian model comparison between different
types of CRSs, including a GWB,
finds the most favoured model to be the 
common uncorrelated red noise described by a power-law with
$A = 5.13_{-2.73}^{+4.20} \times 10^{-15}$ 
and $\gamma = 3.78_{-0.59}^{+0.69}$ (95\% credible regions).
Fixing the spectral index to $\gamma=13/3$ as expected 
from the GWB by circular, inspiralling supermassive black-hole binaries 
results in an amplitude of $A =2.95_{-0.72}^{+0.89} \times 10^{-15}$. 
We implement three different models, \beph{}, \lmoss{} and \egp{}, to 
address possible Solar-system ephemeris (SSE) systematics
and conclude that our results may only marginally depend on these effects.
This work builds on the methods and models from the studies on the EPTA DR1. We show that under the same analysis framework the results remain consistent after the data set extension.

\end{abstract}




\section{Introduction}
\label{sec:intro}

Radio pulsars, and especially radio millisecond pulsars (MSPs), 
have been used as astronomical tools 
to study aspects of fundamental physics with remarkable success,
thanks to their exceptional rotational stability. 
An area of research where MSPs have been particularly useful is gravity 
\citep[e.g.][]{tay1993,ksm+2006,w2014}, especially
by employing the ``pulsar timing'' technique 
\citep[e.g.][]{lk2005}, 
which relies on high-precision measurements of the pulses' 
times-of-arrival (TOAs) being compared to a ``timing model''.
The difference between the measured and the model-predicted TOAs is referred to as the ``timing residuals''. 
Any unmodelled effects will appear in the timing residuals, 
and the timing model is revised and/or extended accordingly.
The timing models have astounding predictive power as it only requires the precise 
modelling of the pulsar's rotation, orbital motion and the signal's propagation in 
space, and not the details of the radiation's physics or emission mechanism.
Pulsar timing observations provided the first evidence for the 
existence of gravitational waves (GWs) \citep{tw1982}, by confirming that the 
measured orbital changes of the binary pulsar PSR~B1913+16 match those predicted 
by the theory of General Relativity (GR) due to the system's energy loss
through the emission of GWs. 

MSPs have been proposed as a tool for the direct 
detection of GWs at nHz frequencies \citep{saz1978,detw79}. 
The experiment is based on systematically observing 
an ensemble of MSPs at many 
sky positions, a configuration called a ``Pulsar Timing Array'' 
\citep[PTA; ][]{fb1990}.
One of the physically motivated sources of GWs in the nHz band
are inspiralling supermassive black-hole binaries (SMBHBs). 
The incoherent superposition of a large number of 
unresolved SMBHB GW signals forms a GW 
background (GWB) that may be detected with a PTA 
\citep[][]{rr1995,jb2003,wl2003,ses+2004}.
In addition, GWBs from cosmic strings 
\citep{kib1976,sbs2012} or a cosmological relic GWB 
from the inflationary era \citep[see][]{gri2005}
have also been proposed as PTA target signals. 
As such, PTAs can provide direct observational 
constraints for large-scale structure and 
cosmological models.
Various theories of gravity, including GR, 
predict that the propagation of GWs cause distortions 
in the space-time metric, 
with specific polarization modes \citep[e.g][]{ew1975,ell+1973,ell1973}. 
GWs propagating in the vicinity of the Earth and the pulsars 
induce spatially correlated variations in the TOAs 
over the timescale of several decades. 

The basic idea of searching for the GWB with PTAs 
is to look for this common red signal (CRS) with the characteristic spatial correlation in the array of MSPs \citep*[see][]{hd1983,ljp2008}.
While PTA data are also being used for other
applications, such as probing Solar-system planetary parameters \citep[e.g.][]{cgl+2018}
establishing pulsar-based timescales \citep[e.g.][]{hgc+2020} and measuring local clock instabilities \citep{llc+2020},
the nHz GW search remains the primary objective of PTA efforts.

After the first experimental efforts to establish a PTA pioneered by Donald
Backer and collaborators (e.g.~\citealt{rom88,fb1990,bac95}), collaborations
with European partners to also utilize 
the Effelsberg 100-m radio telescope (EFF) 
and the Nan\c{c}ay Radio Telescope (NRT) 
started corresponding timing efforts. 
Together with regular timing efforts with the Lovell Telescope (LT) and the Westerbork Synthesis Radio Telescope (WSRT), 
this laid the foundation for uninterrupted 
PTA data sets spanning now up to 24 years for a number of sources. 
Inspired by the formation of the Parkes Pulsar Timing Array (PPTA)
\citep{man06,hob2013}, the European Pulsar Timing Array 
was officially established in January 2006 
\citep{skldj06,jsk+08}
as a collaboration of European radio observatories and research institutions 
working towards the direct detection of nHz GWs.
Apart from the PPTA, the EPTA has also been working alongside
the North-American Nanohertz Observatory for Gravitational Waves 
(NANOGrav, \citealt{jenet09,abb+2015a}).
The three groups collaborate
under the International Pulsar Timing Array (IPTA, \citealt{vlh+2016,pdd+2019}),
with combined data sets and shared resources and expertise. The advent
of further sensitive telescopes able to perform precision pulsar timing
observations, especially the Giant Metrewave Radio Telescope (GMRT, 
\citealt{swarup90}) in India, which recently formally joined the IPTA, and more recently 
the Five Hundred Meter Spherical Telescope (FAST, \citealt{jyg+19}) in China, 
and the MeerKAT telescope \citep{camilo2018} in South Africa with the MeerTIME program \citep{bja+2020}, adds important capabilities to the IPTA.

With continuous improvements of the hardware used,
the current EPTA data set (henceforth Data Release 2; DR2) 
extends the previous data set \citep[DR1;][]{dcl+2016} by up to 7 years.
From early on, coherent-dedispersion data acquisition 
hardware has been in place at some of the telescopes 
(e.g. the Effelsberg Berkeley Pulsar Processor 
(EBPP), the Berkeley Orl\'{e}ans Nan\c{c}ay Instrumentation (BON), 
or PuMa II at the WSRT, see Section~\ref{sec:data}). 
Additionally, the Large European Array of Pulsars (LEAP,
\citealt{ks10,bjk+16}) efforts combine the EPTA telescopes in tied-array mode
to form a 194-m equivalent dish
for MSP timing with monthly cadence for the last 8--10 years. As a result, the
EPTA has effectively operated as a ``mini-IPTA'', combining data sets from six
different telescopes. Consequently, the combination of data sets is complex, 
but the availability to cross check the data with multiple overlapping data set helps enormously in identifying and solving instrumental problems.

In the EPTA DR1 GWB upper limit analysis, \cite{ltm+2015} (henceforth LTM15) applied a methodology to probe a number of physically motivated CRSs in the data set and compare their probability against the GWB with the Hellings-Downs (HD) correlation \citep{hd1983}. 
This methodology was created because there was some modest evidence 
in this first data combination of a common signal
and it was recognised that this would likely require
careful examination in a future extended data set.
In this work, we analyze the same 
six pulsars used in EPTA DR1 GW analyses 
\citep{ltm+2015,tmg+2015,cll+2016,bps+2016}, 
and present updated results using the EPTA DR2 on the properties of a CRS in the data. 
Building upon our DR1 methods we include the same noise terms for the pulsar properties as well as the pulse propagation through the ionized interstellar medium (IISM), which has been the standard first-order analysis 
approach in pulsar timing for a number of years.
While we are working on further optimization in modelling 
and analysis methods, which could potentially improve our results,
conducting the analysis here in the same model framework as in DR1
allows us to get a direct comparison and measure of the improvement 
achieved with the new data.
In order to increase confidence in the results, we make use of multiple independently developed analysis codes, some of which are used for the first time, for the parameter estimation, model selection 
and modelling of Solar-system dynamics.
As a first-order examination of the stationarity of the CRS
we also analyze the evolution of this signal from DR1 to DR2.
This work is the first in a series of papers which will present results from multiple types of analysis using a larger number of pulsars with the updated EPTA DR2 in order to fully establish if the CRS persists and what type of spatial correlation it has.

The rest of the paper is organized as follows: 
In Section~\ref{sec:data} we present the properties of the data used in this study.
Section~\ref{sec:framework} provides an overview of the modelling
and analysis framework. Results from the single-pulsar analysis 
are presented in Section~\ref{sec:spna}. 
Section~\ref{sec:GWB} shows the results from a search for a CRS with
simultaneous analysis of the signal's angular correlations and 
Bayesian model selection
between different possible types of CRSs to gauge which is more 
supported by the data. The most supported model is further
investigated in Section~\ref{sec:test_curn}, 
where we investigate effects from choices in modelling the CRS spectrum, 
the pulsars' contributions to the common signal, 
the time-stationarity of the CRS and the effects of the SSE.
We discuss the results in the framework of 
the GWB and compare these to results from the literature in 
Section~\ref{sec:disc_compare} and present our conclusions
in Section~\ref{sec:conclude}.

\section{Data}
\label{sec:data}

\begin{figure*}
\centering
\hspace*{-2cm}
\vspace*{-0.7cm}
\includegraphics[scale=0.60]{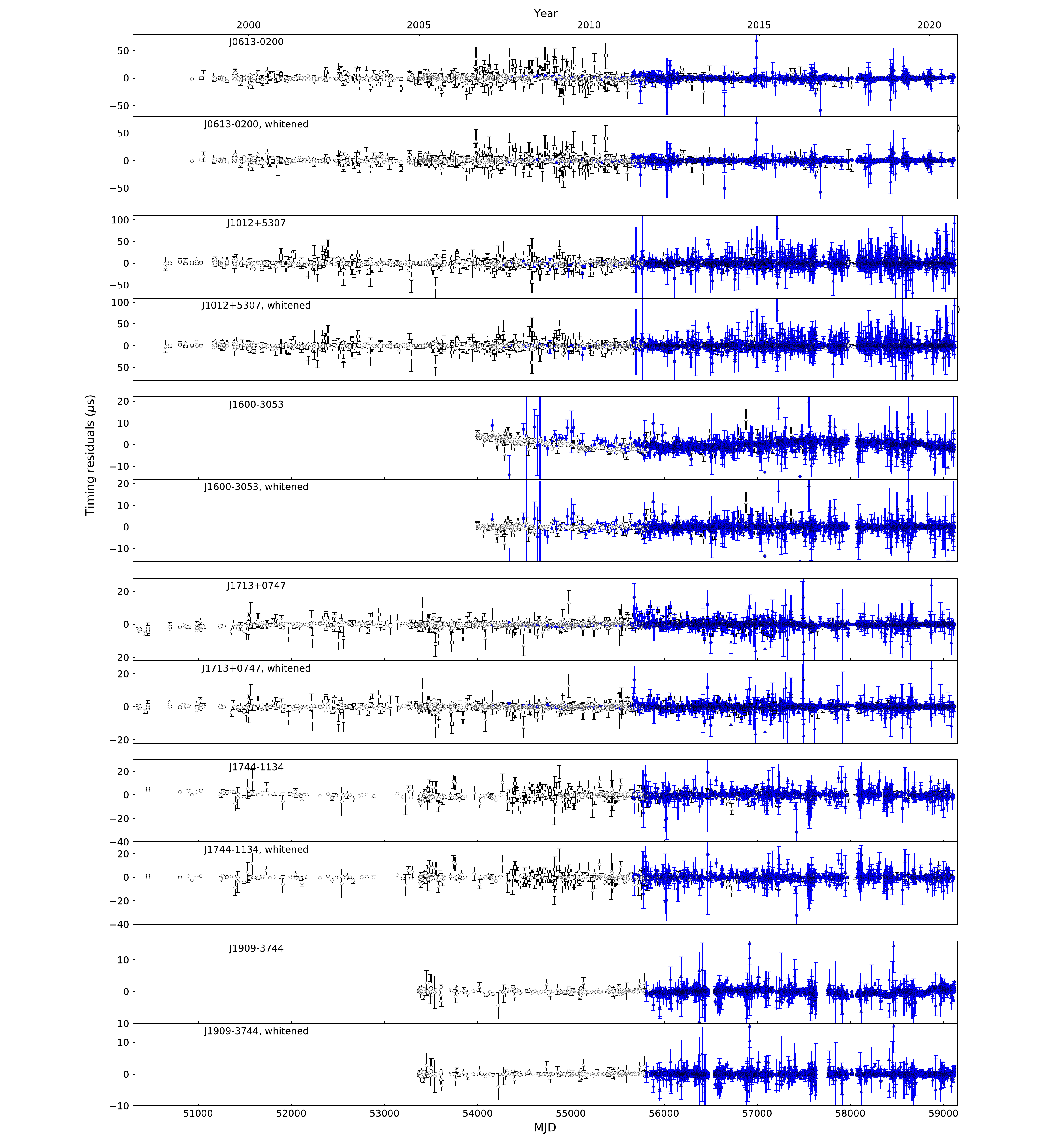}
\caption{Timing residuals of the six pulsars used in this paper. For each pulsar, the residuals before and after subtraction of DM and red noise are shown. The squares, circles and triangles represent P-band, L-band and S/C-band observations, respectively (see Table~\ref{tab:source} for information on the observing frequency bands). The blue/filled and black/unfilled symbols indicate the new backend data and DR1 TOAs, respectively.} 
\label{fig:timing}
\end{figure*}

\begin{table*}
	\begin{center}
		\caption[]{Overview of the six-pulsar data set used in this paper's analysis. 
			The columns represent the name of the pulsar, the coverage of frequency band, the number of TOAs, the timespan $T$ of the data set, the median $\sigma_{\rm TOA}$ in each frequency band, the timing residual rms, the rms after subtraction of DM noise, and the rms after subtraction of both red and DM noise (whitened). The corresponding frequency ranges for the P, L, S and C bands are 0.3--1.0, 1.0--2.0, 2.0--4.0, and 4.0--8.0\,GHz, respectively.}
		\begin{tabular}{ccccccccccc}
			Pulsar name & Band & $N_{\rm TOA}$ &$T$ &\multicolumn{4}{c}{Median $\sigma_{\rm TOA}$ ($\mu$s)}   & rms & rms, DM removed & rms, whitened \\
			\cline{5-8} 
			& & & (year) &P &L &S &C & ($\mu$s) & ($\mu$s) & ($\mu$s) \\
			\hline \noalign {\smallskip}
			J0613$-$0200 &L, S, C & 3022    & 22.4&--     &1.527  & 7.113 & 7.400 &1.415 &1.281 &1.168 \\
			J1012+5307 &P, L, S, C & 5837   & 23.2&4.800  &1.705  & 7.859 & 4.756 &1.393 &1.326 &1.233\\
			J1600$-$3053 &L, S & 3345       & 14.0&--     & 0.475 & 1.762 &--     &1.376 &0.766 &0.439 \\
			J1713+0747 &P, L, S, C & 5052   & 24.0& 1.329 & 0.308 & 0.696 & 0.703 &0.415 &0.355 &0.311   \\
			J1744$-$1134 &P, L, S, C & 1980 & 23.7& 3.700 & 0.912 & 3.185 & 1.046 &0.898 &0.736 &0.653 \\
			J1909$-$3744 &L, S & 2817       & 15.7&--     &0.268  & 0.667 &--     &0.504 &0.424 &0.228 \\
			\hline \noalign {\smallskip}
			\label{tab:source}
		\end{tabular}
	\end{center}
\end{table*}

The EPTA has continued to monitor approximately 50 pulsars 
with high observing cadence since 
the first data release \citep{dcl+2016}, 
using the five European radio telescopes
both in single-dish and LEAP modes.
High-sensitivity TOAs 
produced from coordinated monthly LEAP observations 
are included for the first time in the EPTA data set.
Current observations utilize the new generation of data recording systems at all 
telescopes, primarily using the Re-configurable Open Architecture Computing Hardware
(ROACH) FPGA board developed by the 
CASPER group\footnote{http://casper.berkeley.edu/}, which allows for coherent dedispersion \citep{vb2011}.
At EFF, the observations were conducted mainly at three frequency bands centered 
at 1347, 2627 and 4850\,MHz, and the data were recorded with the `PSRIX' backend with a
bandwidth of 200\,MHz \citep{lkg+2016}. The LT observations 
were carried out at a central frequency of 1532\,MHz, with data recorded using the 
`ROACH' pulsar backend with a bandwidth of 400\,MHz \citep{bjs+2016}. 
At the NRT, 
observations were performed centered at 1484 and 2539\,MHz, and the data were recorded 
with the `NUPPI' backend with a bandwidth of 512\,MHz \citep{ctg+13}.
At the WSRT, observations 
were made at 350, 1380 and 2200\,MHz, and the data recording was performed with the 
'PuMa II' backend \citep{ksv08}. 
The Sardinia Radio Telescope (SRT) is 
the latest telescope addition to the EPTA and now effectively 
participates in EPTA combined data set.
The majority of the SRT 
observations were made as part of LEAP sessions.
Due to the small number of the SRT single-telescope observations
the corresponding TOAs were not included in this data set.
The SRT observations were conducted 
at 1396\,MHz with the data recorded using the `ROACH1' 
backend using a bandwidth of 128\,MHz. 
The LEAP observations were performed at 1396\,MHz 
using the same backends but with a recording bandwidth of 128\,MHz at each telescope.

The six DR1 priority pulsars used in this paper, are: 
PSRs~J0613$-$0200, J1012+5307, J1600$-$3053, J1713+0747, J1744$-$1134 and J1909$-$3744. 
Each pulsar is regularly observed by all the EPTA telescopes, 
except for J1909$-$3744 which given its sky location 
has so far only been monitored by NRT and the SRT;
however, as noted above, single-telescope SRT data are not 
included in the current data set.
The data from the new backends
collected at each telescope were processed using the \psrx{} 
software package \citep{hvm04}, to carry out calibration 
and radio-frequency interference mitigation. 
Then, for each observation epoch, 
an ``integrated profile'' was formed by averaging the 
data in time and frequency. For data from EFF, WSRT and LEAP, 
the frequency averaging was 
performed over the entire band. 
For data from LT and NRT, the averaging was done in two and four 
sub-bands, respectively, to accommodate their larger bandwidth. 
The TOAs of these profiles were then calculated 
using the canonical template-matching method \citep{tay1992}.

We note that for about three years,
observations were made simultaneously both
with the older backends (as they appear in DR1)
and the new ones
with the corresponding TOAs from the new backends.\footnote{These periods of observations with simultaneous data recordings 
using older and newer instruments are intentionally carried out at the observatories
in order to confirm the good performance of the new instruments during commissioning,
and to accurately measure the necessary phase offset between the two data sets,
as required when creating the combined data set.}
In this work, we replace all those DR1 TOAs
We also exclude single-telescope data from epochs that were used to create the corresponding LEAP TOAs.
As seen in Fig.~\ref{fig:timing}, although the timespan extension from
DR1 is $\sim7$ years, the effective improvement due to the new-generation data is $\gtrsim10$ years.

A summary of the combined data set can be found in 
Table~\ref{tab:source}. 
We refer to \cite{lkg+2016}, \cite{bjk+16}, \cite{psb+18}, \cite{lgi+20}, 
and the forthcoming EPTA DR2 paper for more details of the observations and data from each individual telescope.
The full details for the backends, data and TOA extraction for the DR1 data
that are part of this data set can be found in \cite{dcl+2016}.

\section{Analysis Framework}
\label{sec:framework}

In this section, we briefly summarize the established mathematical and algorithmic framework used in PTAs to analyze TOAs.
While the information described in the section
can be found in the literature, we provide this overview
as the methods were progressively created in many publications.
This section can also serve as a quick reference for the planned 
follow-up papers.
The interested reader can study the provided references for 
detailed explanations and formula derivations.

The analysis is divided in two main parts:
(a) the single-pulsar analysis, which provides the pulsar timing parameters
and stochastic noise parameters for each MSP, and
(b) the CRS analysis, which describes the methods used in 
the search for CRSs, including the GWB, 
in the TOAs of all MSPs and the investigation of their spatial correlations.

\subsection{Single-pulsar Timing and Noise Modelling}
\label{sec:theory_spa}

Our search for common signals
between pulsars is preceded by single-pulsar analyses.
This process provides pulsar models that comprise of 
the timing and noise parameters.
The former induce deterministic signals,
while the latter induce stochastic signals.
In the case of GW searches with PTAs,
the pulsar noise analysis is equivalent to the characterization 
of a GW-detector noise and is thus a necessary step before the 
search analysis,
as pulsar noise can correlate with GW signals  
and reduce the data's sensitivity \citep[see e.g.][]{cll+2016}.
All sources of noise therefore need to be measured
in order to be decorrelated from the signal of interest,
and to properly evaluate the possibility 
for the existence of any common signals.

Timing parameters are typically measured progressively
as the pulsar data set increases with new data. 
Least-squares linear fits
are very effective in producing phase-connected timing models 
(i.e. models that account for every pulsar rotation).
The timing analysis in all cases was performed using
the pulsar timing package \ttwo{} \citep{hem2006}. 
It is used to first 
derive the basic phase-connected timing solution which uses a linear approximation for 
the pulsar timing model, i.e. assuming that the timing-model parameters may 
only have small, linear deviations from the true values.
In practice, this means that 
on every update of the model
we assume that there are only 
linear deviations from the pre-fit values of the 
timing parameters to their post-fit values. 
The topocentric TOAs (i.e. TOAs at the observatory) 
were mapped to the Solar-system barycentre to form the barycentric TOAs
using the Solar-system ephemeris (SSE) DE438 from the 
Jet Propulsion Laboratory \citep{fp2018}.
The TOAs were referred to the terrestrial timescale
BIPM2019, provided by the Bureau International des Poids et Mesures\footnote{\url{www.bipm.org}}.

While the least-squares linear approximation timing model
is only valid if the timing data are white-noise dominated, 
pulsar data often contain other types 
of non-white (i.e. time-correlated) noise processes.
In the presence of such correlated noise processes, even for phase-connected
timing models, the values and uncertainties of the timing parameters 
can be biased \citep[e.g.][]{chc+2011}. 
A full analysis of pulsar timing data 
with the intention of accurate pulsar 
parameter measurements is therefore better achieved via a 
simultaneous fit of the timing and noise models. 
In this work we measure the stochastic noise
using the linear approximation to the timing model 
by analytically marginalizing the likelihood over 
the timing parameters, as we show below. This accounts for possible covariances between the timing and noise parameters and has consequently been used in standard PTA analyses.
We note that a more precise timing model can be
derived iteratively with \ttwo{} 
via a generalized least-squares fit, 
using the covariance matrix 
of a noise model \citep[see e.g.][]{chc+2011}, 
such as the maximum likelihood model from the 
Bayesian noise analysis; this approach 
was used to produce the timing residuals
presented in Fig~\ref{fig:timing}.

In this study, we use the same pulsar noise models
as in the EPTA DR1 analysis \citep{dcl+2016,cll+2016}. 
The model uses two white-noise 
parameters to properly re-scale the TOA uncertainties, 
an achromatic red noise
component, and a chromatic dispersion-measure (DM) noise component. 
We discuss these noise components, in brief.

The noise analysis is conducted using a Bayesian framework,
following the same general approach as in the EPTA DR1 \citep{cll+2016},
and uses two independent analysis codes to cross-check for consistency in the results
(see Section~\ref{sec:spna_codes} for details).
We present an overview of the mathematical configuration and provide basic details for the calculations
of the noise covariance matrices, at the level where the used
analysis codes follow the same principles, 
even though differences can appear in the 
exact programming implementations of these principles.  

For a single-pulsar timing problem, where we have $n$ TOAs, 
the likelihood function was first introduced in \cite{vlm+2009} as,
\begin{equation}
\label{eq:psrlik}
L_{\rm{PSR}}=\frac{\e^{-\frac{1}{2}(\res_{\rm post})^{\tr}\C^{-1}(\res_{\rm post})}}{\sqrt{(2\rmpi)^n|\C|}}\,.
\end{equation}
We use the subscript $_{\textrm{PSR}}$ to denote that the function 
is for the single-pulsar problem, and the $^{\textsf{T}}$ superscript to denote the matrix transpose.
The likelihood function is derived under the 
assumption of operating in the linear approximation of the timing model,
where the initial values of the timing parameters are close to the
correct ones and only linear deviations are
required when re-fitting the data. In this way,
the (post-fit) timing residuals are estimated
with the use of the design matrix of timing parameters,$\M$,
and the vector $\xi$ of the amplitudes
of the timing-parameter signals, 
as $\res_{\rm post} = \res_{\rm pre} - \M\xi$,
where $\res_{\rm pre}$ are the pre-fit residuals 
(i.e. prior to the new fit of the timing model).
The effects of the additional stochastic parameters are described
via the pulsar's total covariance matrix, $\C$.
Our noise model assumes three 
stochastic noise components in the data:
white, red and DM. 
Each of these noise components is defined by a covariance matrix: $\Cw$, $\Cr$, and $\Cdm$, respectively.
The total covariance matrix is the sum of these,
\begin{equation}
\label{eq:covmat}
\C = \Cw + \Cr + \Cdm \,.
\end{equation}
\cite{vlm+2009} describe a way 
to analytically marginalize
the pulsar timing parameters 
using the reduced likelihood function,
\begin{equation}
\label{eq:psr_reduc_lik_mm}
L'_{\rm{PSR}} = \frac{\e^{-\frac{1}{2}(\res_{\rm pre})^{\tr}\Cda(\res{\rm pre})}}{\sqrt{(2\rmpi)^{n-m}\times|\C|\times|\Mtr\Cinv\M|}}\,,
\end{equation}
where $\Cda=\Cinv-\Cinv\M(\Mtr\Cinv\M)^{-1}\Mtr\Cinv$,
and $m$ is the number of timing parameters that are marginalized.
In this work, the analytical marginalization over the 
timing parameters is performed in one of two
ways, depending on the computing algorithm used
(discussed later). We briefly discuss these methods
in chronological order of their appearance in the literature.

One approach currently used for marginalizing the timing model,
which we refer to as the ``G-matrix'' approach,
is a more numerically stable implementation of the initial formulation.
It uses a reduced likelihood function defined as \citep[][]{vl2013},
\begin{equation}
\label{eq:psr_reduc_lik_gm}
L'_{\rm{PSR,1}} = \frac{\e^{-\frac{1}{2}(\res{\rm pre})^{\tr}\G(\Gtr\C\G)^{-1}\Gtr(\res{\rm pre})}}{\sqrt{(2\rmpi)^{n-m}\times|\Gtr\C\G|}}\,,
\end{equation}
where $\G$ can be derived from the design matrix via a singular value decomposition \citep[see][for details]{vl2013}. As in equation~\eqref{eq:psrlik},
The dimensions of the $\C$ and $\G$ are
$n\times n$ and $n\times(n-m)$, respectively.

The second marginalization method
follows a different route, and we
refer to it as the ``Gaussian-process'' approach \cite[e.g.][]{vv2014,abb+2016}.
The post-fit residuals are now expressed by taking into account effects from the stochastic noise components in addition to the timing parameters via the design matrix.
Further, the timing model effects are added to the total covariance matrix (equation~\eqref{eq:covmat}) 
resulting in $\Cda_2 = \Cw + \T\B\Ttr$.
The matrix $\T$ includes the design matrix in addition to terms related to the stochastic noise and $\B$ is a matrix used to define the 
prior distributions of the timing 
and noise-related parameters (these are frequency-domain 
parameters that sample the noise, 
as discussed later in this section 
(see e.g. equation~\eqref{eqn:fourier}).
In this case, the marginalized likelihood becomes,
\begin{equation}
\label{eq:psr_reduc_lik_gp}
L'_{\rm{PSR,2}} = \frac{\e^{-\frac{1}{2}(\res_{\rm pre})^{\tr}\Cdainv_2(\res_{\rm pre})}}{\sqrt{(2\rmpi)^n\times|\Cda_2|}}\,,
\end{equation}
By setting uniform and infinite priors 
on the timing parameters via $\B$ and using the Woodbury matrix identity \citep{woo1950,hag1989} on $\Cdainv_2$, \cite{vv2014} have recovered an equivalent to the marginalization in equation~\eqref{eq:psr_reduc_lik_mm} 
times a constant from the prior matrix, $\B$,
that acts only as a renormalization in the Bayesian analysis.

To conclude the basic description
of the single-pulsar analysis, 
we discuss the noise components 
and their respective covariance matrices.
In all cases, the stochastic noise components
are modelled as random Gaussian processes.
Detailed information on the noise modelling
and the likelihood function can be found in \cite{vl2013,vv2014,lah+2013,lah+2014}
and references therein,
so we only provide an overview here for completeness.

\subsubsection*{White noise}

The white noise is modelled with the so-called EFAC and EQUAD stochastic parameters. For each observing system,
defined by the telescope receiver and backend system combination, 
we apply an (EFAC, EQUAD) pair to re-scale the initial 
TOA measurement uncertainty, $\sigma$, according to equation,
\begin{equation}
\label{eq:sigma_rescale}
{\hat \sigma}^2 = (\sigma\cdot \textrm{EFAC})^2+\textrm{EQUAD}^2\,.
\end{equation}
EFAC is a multiplicative factor that accounts for possible
errors in the estimation of the formal TOA uncertainty as calculated
during the cross-correlation of the pulse profile with a standard template \citep{tay1992}.
EQUAD is a term added in quadrature 
to account for additional scatter of the TOAs
due to physical effects such as pulsar jitter noise
\citep{em1968}, an effect that appears to be
often relevant in high precision pulsar timing
observations \citep[e.g.][]{ovh+2011,lvk+11,lkl+2012,sod+2014,lcc+2016,lma+2019,pbs+2021}.

The TOA uncertainties are re-scaled such that 
the data conforms to our basic assumption that they
are drawn from a Gaussian parent distribution.
The case of non-Gaussian uncorrelated TOA noise
is discussed theoretically in the literature \citep[see e.g][]{lha+2014,vv2017}, 
and we note that evidence for 
non-Gaussianities in TOAs have recently been
reported for a small number of MSPs \citep{grs+2021}.
Increasing TOA precision may result in the need to
take non-Gaussianities into account in the future.

The white-noise covariance matrix, $\Cw$,
is a diagonal matrix with the 
re-scaled variances of the TOAs (equation~\eqref{eq:sigma_rescale})
as its elements, i.e.,
\begin{equation}
\label{eq:TN_Cr1}
C_{\textsf{W},i,j}={\hat \sigma}_{ij}^2\delta_{ij} \,.
\end{equation}

\subsubsection*{Red and dispersion-measure noise}

The red noise component addresses 
the pulsar's intrinsic low-frequency noise,
achromatic noise, which has mostly been 
associated with pulsar-spin 
irregularities \citep[e.g.][]{cd1985,dmh+1995}.
While MSPs have much lower levels of red noise than young pulsars, 
the effects are 
measurable, especially for long-term data sets and at this high timing precision \citep[e.g][]{vbc+2009,cll+2016,aab+2021,grs+2021}.

DM is defined as the integrated column density of free electrons in the pulsar's line of sight and can vary over time \citep{yhc+2007a}.
In principle, it is possible to have DM measurements for each observation if a sufficient range of observing frequencies is covered for each observation.
In the case of the EPTA data, however, we are dealing with a very long data set where, due to differing circumstances at our participating telescopes over the last 24 years, 
there is a significant level of inhomogeneity 
in the observing frequencies across the data set.
We therefore opt to fully model the temporal DM variations as a combination of deterministic and stochastic processes, as discussed in \cite{lbj+2014}.
The timing model parameters of the DM value at a reference epoch and its first and second time derivative give a first-order deterministic approximation. Delays from stochastic turbulences of the IISM add an additional DM noise component.

In the models used in this work, 
both the red and DM stochastic noise 
components are described as wide-sense stationary signals 
with a single power-law 
spectrum of the form,
\begin{equation}
S\propto A^2f^{-\gamma}\,,
\label{eq:spectrum}
\end{equation}
where $A$ is the spectrum's amplitude and $\gamma$ its spectral index.

The red-noise covariance matrix is calculated via the 
Woodbury identity \citep{woo1950,hag1989}.
We consider the Fourier transforms $\F$ 
of the time-domain red noise signal $\boldsymbol{t}_{\textsf{R}}$,
which has been shown to be a well-performing
approximation in pulsar timing analysis \citep{lah+2013,vv2015}.
We then have,
\begin{equation}
\label{eq:TN_Cr1}
(\Cw+\Cr)^{-1}=\Cw^{-1} - \Cw^{-1}\F\lbrack\Ftr\Cw^{-1}\F+\Psi^{-1} \rbrack^{-1}\Ftr\Cw^{-1} \,.
\end{equation}
In matrix notation, the Fourier transforms are,
\begin{equation}
\boldsymbol{t}_{\textsf{R}} = \F\textit{\textbf{a}} \,.
\label{eqn:fourier}
\end{equation}
The Fourier elements
are defined by a limited number of sine-cosine pairs
$\{\sin(2\rmpi{}tf); \cos(2\rmpi{}tf)\}$
with coefficients \textit{\textbf{a}} 
and corresponding covariance matrix 
$\Psi=\langle{} \textit{a}_{f} \textit{a}_{f'}\rangle$
(where the indices $f$, $f'$ correspond to the different frequencies).

In contrast to the red noise, the DM noise is chromatic and
the model demands that the induced TOA delays are related to the inverse square
of their observing radio frequency, $\nu$,
following the dispersion law of cold plasma \citep[e.g.][]{ll1960}.
The covariance matrix of the 
DM stochastic noise is calculated using the same recipe 
as in the case of the red noise component, with one
change that introduces the dependency of the induced time delay
on the observing frequency \citep{lbj+2014}. 
To achieve this in the frequency-domain noise models used here, 
the sine and cosine Fourier transform elements are multiplied by
the corresponding (in time) elements of a vector $\textsf{\bf F}^{\textrm{DM}}$
with length equal to the number of TOAs 
and elements (LTM15),
\begin{equation}
\label{eq:DMF}
\textsf{F}^{\textrm{DM}}_i=1/K\nu_i^2\,,
\end{equation}
where $K=2.41\times10^{-16}($Hz$^{-2}$cm$^{-3}$pc\ s$^{-1}$)
is the dispersion constant and $\nu_i$ is the observing frequency 
of the i-th TOA.

\subsubsection*{J1713+0747 events}

The TOAs of PSR~J1713+0747 are characterized by two additional structures in the timing residuals in this data set, 
caused by discrete ``events'' affecting the pulsar's signal.
They manifest as a rapid decrease in the timing residuals, followed 
by a recovery occurring on timescales of the order of $\sim100$ days. 
The first of these events has been detected by all PTAs 
\citep[see][]{dfg+2013,kcs+2013,dcl+2016} 
and has been linked to discrete variations in the DM 
parameter that are not strictly caused by the turbulence in the IISM.
The second one has previously been reported in \cite{leg2018,grs+2021} with some evidence indicating a deviation from the DM nature of the event. These events have been physically interpreted 
as being caused by plasma lensing effects, due 
to discrete, under-dense IISM structures \citep{leg2018}.
Alternatively, \cite{grs+2021} claim that the second event 
could be related to a sudden change in the pulsar profile.
As these events are non-stationary, 
they are not accurately accounted for 
when modelling the power spectrum of the 
long-term DM and pulsar red noise.

The first of these events was modelled in the EPTA data using shapelet basis functions
\citep[see][]{dcl+2016} and we follow the same strategy for both events in this work.
A comparison of the single pulsar analyses with and without these two events shows 
significant changes in the parameters of the 
DM stochastic component, but mostly unaffected red noise properties. 
When not including the events, the stochastic DM noise component has a 
larger amplitude which compensates for 
the spectral power located in the DM variations induced by the events. 
However, we do note that quantitative model comparisons 
significantly support the model with the
additional events for PSR~J1713+0737.
We will address these details, including a comparison 
with findings from other PTAs \citep{hst+2020,grs+2021}
in an upcoming paper that 
will focus on more precise noise models for the EPTA DR2 data (Chalumeau et al., in prep.).
We have verified with test runs that modelling or not these events separately does not 
affect the CRS analysis, as the red noise properties of the MSP
(which is the noise that can correlate with a GWB or other achromatic CRSs)
remain unaffected by including or not the parameters for the additional events.
A more detailed discussion can be found in Chalumeau et al. (in prep.).
Therefore, we note that for simplicity in this work the GWB search and all analyses 
regarding the search for common signals discussed in 
Sections~\ref{sec:GWB} and~\ref{sec:test_curn}
use the simpler model that does not include these events.

\subsection{Gravitational-wave background search with Pulsar Timing Arrays}
\label{sec:theory_gwb}

The GWB is a stochastic signal, 
which can be parametrized by a power-law
that describes the GW-frequency dependence of its
characteristic strain, $h_{c}$, a measure of the 
space-time deformation that the GWB induces \citep[see e.g.][]{ maj2000,jhv+2006}.
This dimensionless strain spectrum is defined as,
\begin{center}
\begin{equation}
\label{eq:strain}
h_{c} = A_{\textrm{GWB}}\left(\frac{f}{f_{\textrm{c}}}\right)^{\alpha}\,,
\end{equation}
\end{center}
where $f$ is the GW frequency and $f_{c}$ is a reference frequency
(typically set to 1~yr$^{-1}$), 
$A_{\textrm{GWB}}$ is the GWB strain amplitude at reference frequency 
and $\alpha$ is the spectral index, 
which varies based on the physical origin of the GWB.
Conveniently, we can express the effect of $h_c$ on the
observed pulsar timing residuals, via the one-sided power spectral density
of the GWB induced residuals as,
\begin{center}
\begin{equation}
\label{eq:gwb_S}
S_{\textrm{GWB}} = \frac{A_{\textrm{GWB}}^2}{12\rmpi^2}\left(\frac{f}{f_{c}}\right)^{-\gamma_{\textrm{GWB}}} f_c^{-3}\,.
\end{equation}
\end{center}
The spectral indices of the GWB characteristic strain
and power spectrum are related as $\gamma_{\rm GWB} = 3 - 2\alpha$.
For a GWB from the cosmic SMBHB population,
the value of $\alpha$ depends on the astrophysical 
details of the inspiralling SMBHBs, such as whether
their orbital dynamics are coupled with their stellar
and gaseous environments \citep[see e.g.][]{ses2013,csd2017}.
For the case of circular, GW-driven binaries,
the slope is $\alpha = -2/3$ or $\gamma_{\rm GWB} = 13/3$.

The timing residuals induced by an isotropic GWB are
spatially correlated. The GWB-induced residuals comprise of two terms:
the pulsar and Earth terms, which are caused by the spacetime deformation from the propagation of the GWB at the pulsar and Earth positions, respectively.
Only the Earth terms are coherent across all pulsars,
leaving the pulsar terms to act as noise.
The correlation coefficients 
between the timing residuals of all 
pulsar pairs as a function of the pulsar-pairs angular correlations
is called
the overlap reduction function (ORF) \citep[see][]{flr2009}.
Owing to different GW polarization modes or graviton masses,
GR and alternative theories of gravity predict different ORFs
\citep[see][for an overview]{lee2013}.
In the context of GR, GWs
have a quadrupolar spatial correlation.
As pulsars 
are separated from each other and from the Earth 
by multiple GW wavelengths (short-wavelength approximation)
the ORF is approximated by the 
Hellings-Downs (HD) curve \citep{hd1983}, with terms calculated as,
\begin{center}
	\begin{equation}
	\label{eq:HDcurve}
	\Gamma_{\textrm{GWB}}(\zeta_{IJ})=\frac{3}{2}x_{IJ}\ln{x_{IJ}}-\frac{x_{IJ}}{4}+\frac{1}{2}+\frac{1}{2}\delta{x_{IJ}}\,.
	\end{equation}
\end{center}
Here, the $I,J$ indices represent the different pulsars,  
$\Gamma_{IJ}$, is the predicted correlation coefficient per pulsar pair
with angular separation $\zeta_{IJ}$, $\delta(x_{IJ})$ is the Kronecker delta,
and $x\equiv[1-\cos(\zeta)]/2$. 
Note that due to the isotropic
nature of the GWB, only the angular separation between pulsar pairs is important,
and not the position of each pulsar. 
The measurement of the ORF curve is central in every method
developed for GWB searches and detection,
as PTAs can in fact detect different CRSs other than
the GWB, as we discuss below. 
Therefore, no GWB detection claim
can be made without a statistically significant measurement of the HD curve.
As such, the search for a GWB essentially amounts
to searching for a CRS in the PTA data, determining its spectral properties
and ORF and comparing the results with the theoretical predictions of a HD correlated signal.

For joint analysis of all pulsar data in search 
of a CRS like the GWB, the likelihood function 
can be expressed by generalizing equation~\eqref{eq:psrlik} as,
\begin{equation}
\label{eq:gwb_reduc_lik}
L_{\rm{PTA}}\propto\frac{\e^{-\frac{1}{2}\sum_{I,J,i,j}(\widetilde{\delta{}t}_{I,i})^{\tr}\Ctot^{-1}(\widetilde{\delta{}t}_{J,j})}}{\sqrt{|\Ctot|}}\,.
\end{equation}
We use the subscript $_{\rm{PTA}}$ to indicate that 
the equation is for the multi-pulsar case,
and we use tilted overlines
to indicate the corresponding data and matrices.
Therefore, $\widetilde{\delta{}t}$ is a vector of
all pulsar timing residuals concatenated.
The total covariance
matrix, including the CRS signal is then,
\begin{equation}
\label{eq:covmattot}
\Ctot = \Ctot^{*} + {\textsf{\textbf{C}}_{\textsf{CRS}}}\,.
\end{equation}
where $\Ctot^{*}$ is 
the block diagonal of all pulsar covariance matrices.
For any CRS with ORF $\Gamma_{\textsf{CRS}}(\zeta_{IJ})$,
its covariance matrix elements are calculated as,
\begin{equation}
\label{eq:gwb_C}
C_{\textsf{CRS},I,J,i,j} = \Gamma_{\textsf{CRS}}(\zeta_{IJ})C_{\textsf{CRS}}(i,j)\,,
\end{equation}
where $C_{CRS}(i,j)$ is the CRS-induced time-correlation,
that is calculated as in the case of pulsar red noise (equation~\eqref{eq:TN_Cr1}).
The corresponding reduced likelihood will depend on the 
method chosen to marginalize over all the timing parameters, 
as discussed in Section~\ref{sec:theory_spa}.

It is also possible to assume the ORF and 
derive the corresponding spectral
properties. 
As we will see in Section~\ref{sec:crs_BF}, 
these types of analyses can provide a basis for examining 
how well a given model fits the data 
with respect to another competeting model.
This is a strategy followed when the CRS analysis
cannot result in a clear measurement on the ORF shape.
Here, we briefly discuss the main CRS models other than the GWB
that are of particular interest to PTAs and which we examine separately
in Section~\ref{sec:crs_BF}.
We note that these models were also discussed and investigated
in the framework of the EPTA DR1 GWB upper limit work
presented in LTM15.
A general study and discussion of the effects of these types of 
signals in pulsar timing data and PTA GW searches 
can be found in \cite{thk+2016}.


\subsubsection{Common uncorrelated red noise}
\label{sec:curn}

PTA data may contain a common, but spatially uncorrelated red noise 
(we denote this with CURN).
A CURN reflects the situation where 
the individual pulsar red noise of all pulsars include some
noise component which have very similar spectral properties,
either due to common physical origin or by chance. 
The CURN case, therefore, is not technically a common term,
as it is not induced in every pulsar TOAs from the same extrinsic process.
This would, however, mimic a CRS,
but the calculated ORF of all pulsar pairs 
would be expected to be randomly distributed around zero,
therefore having no spatial correlations.
The CURN is assumed to also 
be described by a red signal with a power-law spectrum,
following equation~\eqref{eq:spectrum}, 
with a unique amplitude and spectral index, $A_{\textrm{CURN}}$ and $\gamma_{\textrm{CURN}}$. 
The covariance matrix $\C$ 
calculated as in the case of pulsar red noise in equation~\eqref{eq:covmat}.
In this case, for each pulsar, $I$, 
the total red noise covariance matrix becomes
$\Cr'(I) = \Cr(I)+\Ccu$. 
In turn, this will affect the 
(block-diagonal) multi-pulsar covariance matrix that we now 
use as the total covariance matrix $\Ctot$ in the 
likelihood function (equation~\eqref{eq:gwb_reduc_lik}).

\subsubsection{Clock-error monopolar signal}
\label{sec:clkn}

One source of possible correlated CRS in 
pulsar TOAs is an error in the terrestrial timescale
to which we refer all TOAs (we denote this with CLK). 
For example, as noted
in Section~\ref{sec:data}, the TOAs in this work were 
referred to the BIPM2019 time-scale. 
As all TOAs are referred to the
same timescale, any imperfections in this timescale
will affect the TOAs of all pulsars in a fully correlated way.
This effect has been extensively discussed in the literature
and it is exploited in order to create pulsar-based
timescales that can be complementary to atomic timescales 
\citep[e.g.][]{gp1991,hcm+2012,hgc+2020}. 

The CLK signal in this work is also modelled as a stochastic red signal
with a power-law spectrum as in previous studies \citep{ltm+2015,cll+2016,hgc+2020}
with parameters  $A_{\textrm{CLK}}$ and $\gamma_{\textrm{CLK}}$.
It has a monopolar correlation, 
with ORF terms,
\begin{equation}
\label{eq:CLKcurve}
\Gamma_{\textsf{CLK}}(\zeta_{IJ})=1, \ \forall (I,J)\,.
\end{equation}

\subsubsection{Ephemeris-error dipolar signal}
\label{sec:eph}

Possible errors in the planetary masses and orbital elements
in the used SSE can induce TOA delays in the PTA
pulsars with dipolar spatial correlations
(we denote this with EPH).
These signals reflect the oscillation of the calculated Solar-system barycentre
position with respect to the true position.  
As in the case a of clock-error signal, this can also be exploited 
in order to provide upper limits or measurements of planetary parameters
using PTA data \citep[see e.g.][]{chm+2010,cgl+2018,gll+2019}.
Unlike the other signals discussed in this Section, the SSE
signals are deterministic signals associated with planetary orbits.
The ORF is defined as \citep[see e.g.][Appendix A]{thk+2016},
\begin{center}
	\begin{equation}
	\label{eq:EPHcurve}
	\Gamma_{\textsf{EPH}}(\zeta_{IJ})\propto\cos(\zeta_{IJ})\,.
	\end{equation}
\end{center}
%

\subsection{Bayesian Analysis}
\subsubsection{Parameter Estimation}
\label{sec:spna_estimation}

We now remind the reader of the basics of 
parameter estimation using Bayesian inference, 
as this is central to this paper's work.
For in-depth descriptions of these methods, the interested reader
can find information in technical textbooks, such as \cite{gre2005}.
The same parameter estimation methods
apply both to the single-pulsar and the common-signal
analyses.

Bayes' theorem is the central equation of the analysis
and is expressed as,
\begin{equation}
\label{eq:bayesT}
Pr(\Theta|\mathcal{D},H)=\frac{Pr(\mathcal{D}|\Theta,H)Pr(\Theta|H)}{Pr(\mathcal{D}|H)}\Leftrightarrow 
P(\Theta)=\frac{L(\Theta)p(\Theta)}{Z}\,.
\end{equation}
In this equation, which we have written in the extended
and a simplified version, $\mathcal{D}$ denotes our data,
$H$ denotes the hypothesis (i.e. the model),
and $\Theta$ denotes the model parameters.
$Pr(\Theta|\mathcal{D},H)$, is the
probability of the parameter, given the data and the model
and is the posterior probability distribution.
This is the distribution we are interested in estimating
and we denote henceforth as $P(\Theta)$.
$Pr(\mathcal{D}|\Theta,H)$ is the likelihood function, henceforth denoted as $L(\theta)$.

Bayesian inference inherently relies on the assumption of what 
our prior information (or belief) 
is for the probability distribution of the parameters.
This information is encoded in $Pr(\Theta|H)$, the probability
of the parameter given the hypothesis/model, 
which is known as the prior probability distribution (or simply the prior)
and henceforth denoted as $p(\Theta)$.
The inclusion of priors is integral in Bayesian inference and the choice
of prior distributions has a central effect on the outcome of the inference,
unless we are in the high signal-to-noise (S/N) regime, at which point 
the posterior distribution becomes insensitive to the choice of prior. 

The final term in Bayes' theorem,
$Pr(\mathcal{D}|H)$, is the probability of the data given the model. 
This is known
as the evidence or marginal likelihood, henceforth denoted as $Z$. 
Note that this term is independent
of the model parameters, $\Theta$. Therefore,
it is not necessary for parameter estimation, 
but it is used for comparing the relative probabilities between models
(and will be implemented in work presented in Section~\ref{sec:crs_BF}).

In Bayesian inference we define the
likelihood function and the prior distribution,
and use the data to update the posterior distribution
of the model parameters.
For most practical problems, 
this requires Monte-Carlo (MC) sampling algorithms.
For the single-pulsar noise parameter estimation discussed
in this paper, we only use pre-determined models 
as discussed in Section~\ref{sec:theory_spa},
therefore we are not concerned with accurate estimations of 
the evidence, $Z$. Finding a more optimal noise model
for EPTA pulsar data will be the
topic of a separate paper (Chalumeau et al., in prep.).
Parameter estimation therefore is 
performed by sampling from the unnormalized
posterior distribution and can be expressed as,

\begin{equation}
\label{eq:bayes_par_est}
 P(\Theta)\propto{}L(\Theta)p(\Theta)\,.
\end{equation}

\subsubsection{Model Selection with Bayes Factors}
\label{sec:modelselect}

In Section~\ref{sec:crs_BF}, we will be
comparing models of different CRSs
in order to evaluate the relative ability of the models
to sufficiently describe the data.
This is described by the posterior odds ratio
of the two analyses.
We express this via the Bayes Factor, $\BF$ as,
\begin{equation}
\label{eq:postodds}
R = \frac{Z_1}{Z_0}\frac{p_1}{p_0} = \BF_{10}\frac{p_1}{p_0} \,.
\end{equation}
We use the numerical subscripts to denote the model,
where 0 denotes the {\it null hypothesis}, which is the simpler
of the two models in the case of nested models.
We see that in contrast to the case of Bayesian parameter estimation,
for model selection we need a
reliable calculation of the Bayesian evidence, $Z$,
which is the likelihood integrated over the prior distribution
of dimensionality $n$, i.e.,
\begin{equation}
\label{eq:evidence}
Z = \int L(\Theta)p(\Theta)\,d^{n}\Theta.
\end{equation}
The Bayes Factor can be calculated directly from the evidences of the two models $Z_1$ and $Z_0$. 
The ratio of the samples in each model is equal to $\BF$, 
under the assumption that the prior probabilities
for the two compared models are equal.
Alternatively, both models can be compared directly against each other in a hypermodel structure, 
which allows the sampler to decide which of them is more likely \citep{hhh+2016}.

In this context, the value of $\BF$ is used
as a metric for the model selection problem.
The difficulty lies in the numerical interpretation
of the Bayes Factor and empirical guidelines
exist based on studies of multiple problems.
In other words, the $\BF$ values
are not calibrated on a scale to provide
a detection significance for this problem
and can fluctuate between 
different analysis runs.
In this paper we will comment on our
results according to the interpretation categories 
listed in tables from \cite{kr1995}, 
noting that in any case these singular $\BF$ values
only serve the purpose of discussing evidence in 
favour or not of one model with respect to another.
As noted in \cite{kr1995}: 
``these categories are not a calibration 
of the Bayes Factor but rather a rough descriptive 
statement about standards of evidence in scientific investigation''.

\begin{table}
    \caption{The prior ranges for the analyses using 
    single power-law spectra models with parameters Amplitude, $A$, and spectral index, $\gamma$. 
    We use the
    subscripts RN, DM and CRS to denote the
    red noise, DM noise and CRS for any type
    of common noise.
    We also include the pulsar white-noise parameters
    EFAC and EQUAD.
    Uniform and log-Uniform refer to flat priors and priors that are flat in the $\log_{10}$ space, respectively}
    \def\arraystretch{1.5}
    \begin{tabular}{l|c|c}
    \hline
    \hline
        Parameter & Prior Type & Range\\
    \hline
        $A_{\rm{RN}}$, $A_{\rm{DM}}$, $A_{\rm{CRS}}$ & log-Uniform & $[10^{-18} - 10^{-10}]$ \\
        $\gamma_{\rm{RN}}$,~$\gamma_{\rm{DM}}$,~$\gamma_{\rm{CRS}}$,~$\delta_{\rm{CRS}}$ & Uniform & $[0 - 7]$ \\
        EFACs & Uniform & $[0.1 - 5]$ \\
        EQUADs & log-Uniform & $[10^{-9} - 10^{-5}]$ \\
    \hline
    \hline
    \end{tabular}
    \label{tab:priors}
\end{table}

\begin{table*}
\caption{Results of single-pulsar noise analysis for \eprise{} (EP) and \tn{} (TN). The table shows the median values from the one-dimensional marginalized posterior distribution. The uncertainties
are calculated such that 95\% of the area under the one-dimensional marginalized posterior distribution of the parameter
is symmetrically distributed around the median. 
\tn{} analysis for the pulsars noted with $^{*}$ was performed using the 
\pchord{} sampler.}
\begin{center}
\def\arraystretch{1.5}
\begin{tabular*}{0.88\textwidth}{c||cc|cc||cc|cc}
\hline
 & \multicolumn{2}{c|}{$\log_{10} A_{\rm{RN}}$} & \multicolumn{2}{c||}{$\gamma_{\rm{RN}}$} & \multicolumn{2}{c|}{$\log_{10} A_{\rm{DM}}$} & \multicolumn{2}{c}{$\gamma_{\rm{DM}}$} \\
Pulsar & EP & TN &  EP & TN &  EP & TN &  EP & TN \\
\hline \hline
J0613$-$0200    & $-14.93_{-1.1}^{+1.11}$ & $-14.94_{-1.0}^{+1.05}$ & $5.09_{-2.08}^{+1.77}$ & $5.12_{-1.93}^{+1.68}$ & $-12.0_{-0.67}^{+0.33}$ & $-12.02_{-0.49}^{+0.31}$ &  $2.55_{-1.07}^{+1.68}$ & $2.59_{-0.95}^{+1.28}$ \\
\hline
J1012+5307$^{*}$    & $-13.13_{-0.17}^{+0.17}$ & $-13.13_{-0.2}^{+0.19}$ & $1.68_{-0.7}^{+0.71}$ & $1.68_{-0.81}^{+0.84}$ & $-11.72_{-0.11}^{+0.1}$ & $-11.73_{-0.12}^{+0.12}$ & $1.21_{-0.39}^{+0.45}$ & $1.2_{-0.42}^{+0.48}$ \\
\hline
J1600$-$3053    & $-14.02_{-1.08}^{+0.51}$ & $-14.01_{-1.13}^{+0.52}$ & $3.46_{-1.49}^{+2.45}$ & $3.45_{-1.43}^{+2.4}$ & $-11.46_{-0.07}^{+0.08}$ & $-11.46_{-0.07}^{+0.07}$ & $2.14_{-0.23}^{+0.26}$ & $2.15_{-0.22}^{+0.24}$ \\
\hline
J1713+0747$^{*}$    & $-14.18_{-0.44}^{+0.35}$ & $-14.17_{-0.52}^{+0.41}$ & $3.37_{-0.98}^{+1.14}$ & $3.35_{-1.13}^{+1.38}$ & $-11.84_{-0.09}^{+0.09}$ & $-11.84_{-0.09}^{+0.1}$ & $1.44_{-0.41}^{+0.43}$ & $1.44_{-0.46}^{+0.51}$ \\
\hline
J1744$-$1134    & $-15.23_{-1.17}^{+1.28}$ & $-15.41_{-0.9}^{+1.19}$ & $5.34_{-2.14}^{+1.58}$ & $5.59_{-1.9}^{+1.27}$ & $-11.69_{-0.11}^{+0.1}$ & $-11.7_{-0.08}^{+0.08}$ & $1.24_{-0.5}^{+0.5}$ & $1.24_{-0.35}^{+0.36}$ \\
\hline
J1909$-$3744    & $-14.57_{-0.77}^{+0.6}$ & $-14.57_{-0.78}^{+0.63}$ & $4.47_{-1.55}^{+1.98}$ & $4.46_{-1.59}^{+1.99}$ & $-12.06_{-0.15}^{+0.11}$ & $-12.06_{-0.16}^{+0.11}$ & $2.03_{-0.44}^{+0.58}$ & $2.03_{-0.43}^{+0.63}$ \\
\hline
\end{tabular*}
\label{tab:spna_dr2_params}
\end{center}
\end{table*}

\section{Single-pulsar Analysis and Results}
\label{sec:spna}

In this section, we discuss
the algorithms and settings used for the single-pulsar analysis and present the results.

\subsection{Analysis algorithms}
\label{sec:spna_codes}

The two algorithms used for pulsar noise analysis 
are \eprise{} \citep{enterprise} and \tn{} \citep{lah+2014}. 
Both packages use Bayesian inference but have been developed independently.
They use different approaches to marginalize the timing-model parameters
and additionally, each code uses different Monte-Carlo samplers
implementing different sampling methods.

The first analysis was performed 
using \eprise{} with \ptmc{} \citep{ptmcmc}, a parallel-tempering 
Markov-Chain Monte-Carlo sampler, that has been designed for a high 
dimensional parameter space. 
To verify the analysis,
we repeated the \eprise{} analysis with a different (nested) sampler, {\tt DYNESTY} \citep{dynesty}. 
Since the results are consistent, only those obtained with \ptmc{} are reported.
\eprise{} employs the ``Gaussian-process'' approach to the marginalization
of the timing parameters (equation~\eqref{eq:psr_reduc_lik_gp}).

The second analysis was performed using \tn{}, 
a Bayesian pulsar analysis package used as a \ttwo{} plugin
which uses a nested-sampling approach \citep{ski2004}.
Depending on the problem's dimensionality we 
either use \mnest{} \citep{fhb2009} or \pchord{} \citep{hhl2015} 
as the sampler,
since the latter is more efficient only in cases with a large number of dimensions.
Following suggestions discussed in \cite{lsc+2016} and 
also based on results from new tests, we used \pchord{} in all cases
where the number of parameters was greater than 55.
We mark the cases where \pchord{} was used in the 
tables reporting the results.
\tn{} uses the ``G-matrix''approach 
for the marginalization of the timing parameters (equation~\eqref{eq:psr_reduc_lik_gm}).

\subsection{Analysis Settings and Results}
\label{sec:spna_results}

We first present some details
regarding the noise analysis settings,
which were set to be common between the two 
analyses. \\ 
\\
(i) In principle, the optimal numbers of frequency components of $\mathbf{F}$ from equation~\eqref{eqn:fourier} to describe the red noise and DM noise stochastic power-laws 
can be a free parameter to be estimated (see Chalumeau et al., in prep.).
For this work we choose conservative numbers based on tests to ensure stable constraints on the power-law parameters. We use identical numbers for all six MSPs, the lowest 30 frequency bins for the red noise and 100 for the DM noise.
Frequency binning is linear at frequencies 
$N/T$, with $N=$1,2,3...$n$,
such that $n/T$ is the highest Fourier frequency of the TOA time series.\\ 
\\
(ii) The prior distributions of power-law amplitudes and 
EQUADs are uniform in log-space, which we refer to as log-uniform 
priors. These types of priors are argued to function as
good approximations of non-informative priors for scale-invariant
parameters \citep{gre2005}.
The spectral indices and EFACs have uniform distributions
in linear space, which we refer to as uniform distributions.
Based on tests performed, we have decided on a
given set of prior ranges for all parameters that we 
concluded to be adequate for this data set. The prior
types and ranges are overviewed in Table~\ref{tab:priors}.

\ \\
We conducted the noise analysis with 
\eprise{} and \tn{} for the EPTA
DR2 data set. The results are 
overviewed in Table ~\ref{tab:spna_dr2_params}.
The two analysis codes have produced nearly identical results.
\footnote{We note that the same analyses
were performed on DR1, showing consistent results.
This is important in both increasing our
confidence in the estimated values of the noise parameters, 
as well as in the validity and quality of the TOAs.}

\begin{figure*}
\centering
\includegraphics[width=\textwidth]{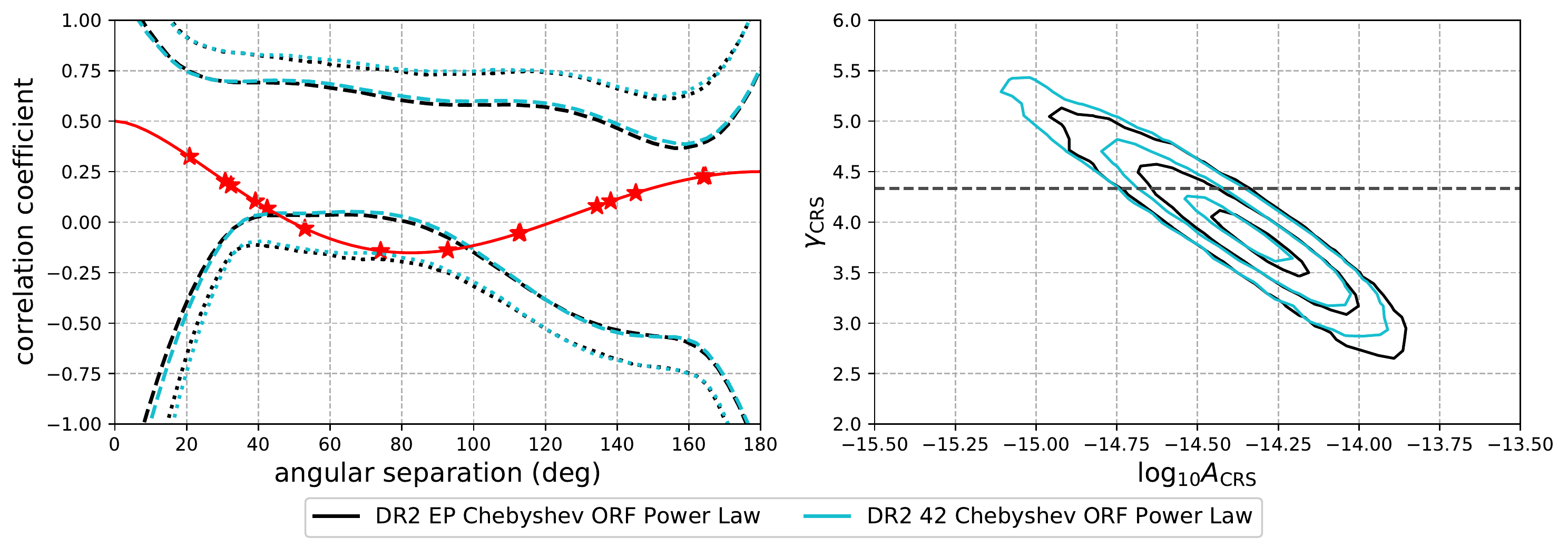}
\caption{\eprise{} 
(EP) and \ftwo{} (42) results from a search for a
CRS modelled with a single-power-law spectrum, 
with simultaneous sampling of the ORF.
The ORF is approximated with
a 4th-order Chebyshev polynomial.
Left: The posterior distribution for the 
angular correlation curve of the CRS, shown
as boundaries of the credible regions. The dashed and 
dotted lines denote
the 95\% and 99.7\% credible regions, respectively.
The theoretical Hellings-Downs curve is overplotted for comparison
as a solid red line. The red stars
denote the angular separations of the pulsars used in this study. 
Right: Two-dimensional posterior distribution of the 
spectral parameters for the single-power-law CRS model with the dashed line indicating the expected $\gamma=13/3$ from a GWB from SMBHBs.}
\label{fig:gen_crs_dr2}
\end{figure*}

\section{Gravitational-wave background search with EPTA data}
\label{sec:GWB}

In this section, 
we present the analysis
and results for a search for a GWB 
in the EPTA data.
We first conduct a search for a power-law spectrum CRS
with simultaneous estimation of the ORF, 
as the most general method to look for the presence 
of a GWB in the data.
Subsequently, we perform spectral parameter estimation
for the physically motivated CRS cases discussed in Section~\ref{sec:theory_gwb}, 
using their predefined ORFs, and employ Bayesian model selection to 
determine which of these models is better supported by the data.

\subsection{Algorithms and analysis settings}
\label{sec:GWBcodes}

For the search of a GWB and other CRSs
in the EPTA data, we once more use two
independently developed algorithms for evaluating the likelihood
and with independent Monte-Carlo (MC) samplers
in order to increase our confidence in the results by having
a cross-check for possible bugs and analysis mistakes.

The first code employed is \eprise{}\footnote{\url{https://github.com/nanograv/enterprise}} with \ptmc{}\footnote{\url{https://github.com/jellis18/PTMCMCSampler}}, 
allowing us to perform a fully integrated pulsar noise and correlated signals analysis within the same analysis suite. We also use \ee{}\footnote{\url{https://github.com/nanograv/enterprise_extensions}} \citep{enterprise_extensions} adding helpful functionality to \eprise{}.
As in the case of the single-pulsar noise analysis, \eprise{}
uses the ``Gaussian-process'' approach to perform the 
marginalization of the timing model.

The second code used is \ftwo{},
which accepts pulsar noise analysis parameters 
from \tn{} results as input for
further analysis. 
Individual modules of \ftwo{} have previously been used in various
publications as they were being developed \citep{cll+2016,ltm+2015,cgl+2018,gll+2019,hgc+2020}
and a unified version is planned to be released in the future.
For MC sampling, \ftwo{} uses either 
\pmnest{}\footnote{\url{https://johannesbuchner.github.io/PyMultiNest/}}
or \pchordlite{}\footnote{\url{https://github.com/PolyChord/PolyChordLite}},
\python{} implementations of \mnest{} and \pchord{}, 
or its own Metropolis-Hastings sampler, depending on the analysis.
To remain consistent with \tn{} \ftwo{} implements the ``G-matrix'' timing parameter marginalization method. The only exception to this
is the analysis in Section~\ref{sec:sseEr} where the timing model
additionally includes SSE planetary parameters, in which case
\ftwo{} also employs the ``Gaussian-process'' approach 
enabling the algorithm to run significantly faster.

In all analyses for common signals in the six
pulsar, their red-noise and DM-noise parameters are 
simultaneously sampled with the CRS. There is a significant 
probability for these pulsar parameters to correlate
with the CRS, especially if the latter is
not detected in the high S/N regime,
either due to its weakness or lack of sufficient pulsar pairs
to disentangle common from non-common signals.
In the single-pulsar analysis, we model the white noise
with two parameters per observing system. Keeping such a
configuration in these CRS analyses would result
in a currently unmanageable number of parameters.
One approach is to fix the EFAC and EQUAD for each observing system and use a ``global EFAC'' parameter per pulsar, 
that acts as a global multiplication factor to regulate 
each pulsar's white noise level.
This has been shown to be a good strategy during the 
EPTA DR1 GWB analysis, as shown also in LTM15,
where in all cases the global EFAC was found to 
be very consistent with unity. This means that the white-noise 
estimation during single-pulsar analysis is very robust.
In the analyses presented in this work, we have verified
the global EFAC values to be $\sim1$ once again, allowing us
to fix the pulsar white-noise parameters from
the single-pulsar analysis without 
significant loss of accuracy in our parameter
estimations. 

The general setup of the correlated search is as follows:\\
\\
(i) For CRS models, 
we use the lowest 30 frequency bins to describe the
single power-law spectrum. The CRS frequency bins are determined 
by a time grid set by the total
timespan of the combined TOA data set.
As we also sample the pulsar red and DM noise parameters simultaneously,
we maintain the same frequency binning for each pulsar as
in the single-pulsar analysis, 
i.e. the lowest 30 and 100 of each pulsar, 
for red and DM noise, respectively. 
As such, although all pulsar red noise components and the CRS
component use 30 frequency bins, 
they all correspond to different frequencies.\\ \\
(ii) The prior distribution of the CRS parameters are noted in Table \ref{tab:priors}.\\ \\
(iii) The majority of the analyses use the DE438 SSE model, without any modification.
Exceptions to this are discussed in Section~\ref{sec:sseEr}.

\subsection{Common-red-signal search and overlap-reduction-function estimation}
\label{sec:crs_results}

We conduct a general search for a
CRS including an estimation of the ORF
using both the \eprise{} and \ftwo{} packages.
We sampled the CRS spectral properties 
using the single power-law spectrum with
amplitude and spectral index $A_{\textrm{crs}}$ and $\gamma_{\textrm{crs}}$,
respectively. The priors are found in Table~\ref{tab:priors}.
The spatial correlation curve of the CRS 
is modelled using a Chebyshev polynomial.
We have introduced this method in LTM15, and we 
simply rewrite the formalism here for easier reading. 
We again use four Chebyshev coefficients ($c_i$, $i\in[1,4]$), therefore the
correlation curve is approximated by,
\begin{equation}
\label{eq:cheby1}
\Gamma(x) \approx c_1 + c_2x + c_3(2x^2-1)+c_4(4x^3-3x)\,,
\end{equation}
where $x=(\zeta_{IJ}-\rmpi/2)/(\rmpi/2)$. 
Chebyshev polynomial priors are flat in the range [-1,1].
The analysis limits the resulting cross-correlations $\Gamma(x) \in [-1,1]$.
In LTM15, we have showed that this approach approximates very well
the direct individual measurements of pulsar-pair cross correlations.
To confidently transition from our DR1 results,
for this analysis we use \ftwo{} with the Metropolis-Hastings sampler as it was
done in LTM15 and compare the results 
with those of the \eprise{} analysis.

Figure \ref{fig:gen_crs_dr2} shows 
the results of this general CRS search.
The analyses clearly recover a common signal,
with the two pipelines providing very consistent results.
The left panel shows the estimation of the 
ORF curve and the right panel shows the posterior distributions
for the spectral parameters.
The one-dimensional posterior distributions
of the spectral parameters are
$\log_{10}A_{\textrm{CRS}}=-14.32_{-0.39}^{+0.31}$
and $\gamma_\textrm{CRS}=3.83_{-0.72}^{+0.82}$,
where we denote the median and uncertainties at
the 95\% credible region (see also Table \ref{tab:pl_js}). 
The ORF figure shows the 95\% and 99.7\% credible regions. The results are compatible with the EPTA DR1 GWB analysis from LTM15.
The boundary encompasses the probability for ORF to be the HD curve,
however other possibilities remain.
We therefore examine in the next section the level of support the 
data provides to the physically motivated CRS signals
discussed in Section~\ref{sec:theory_gwb}.

\begin{table}
\caption{
Results from model selection analysis, in
logarithmic (base 10) Bayes Factors ($\log_{10} {\rm  BF}$), for different CRS models, and with fixed SSE (DE438).
The model-components acronyms are:
(i) PSRN = individual Pulsar noise only, (ii) CURN = Common uncorrelated red noise,
(iii) GWB = isotropic GWB with quadrupolar, Hellings-Downs, angular correlation, 
(iv) CLK = common signal with monopolar spatial correlation, as expected from a clock error,
(v) EPH = common signal with dipolar spatial correlation, as expected from SSE errors. PSRN has no $\log_{10} {\rm  BF}$ values as it serves as the base model.
See Sections \ref{sec:theory_gwb} - \ref{sec:eph} for the discussion on these models.
}
\begin{center}
\def\arraystretch{1.5}
\begin{tabular*}{0.9\linewidth}{c|c||cc}
\hline
 & & \multicolumn{2}{c}{$\log_{10} {\rm  BF}$} \\
ID & Model & \eprise{} & \ftwo{} \\
\hline \hline
0 & PSRN                 & -- & -- \\
\hline
1 & PSRN + CURN          & $3.8$ & 3.6 \\
\hline
2 & PSRN + GWB           & $3.4$ & 3.2 \\
\hline
3 & PSRN + CLK           & $0.6$ & 0.8 \\
\hline
4 & PSRN + EPH           & $2.1$ & 2.1 \\
\hline
5 & PSRN + CURN + GWB    & $3.6$ & 3.7 \\
\hline
6 & PSRN + CURN + CLK    & $3.7$ & 3.4 \\
\hline
7 & PSRN + CURN + EPH    & $3.7$ & 3.4 \\
\hline
\end{tabular*}
\label{tab:bf_dr2}
\end{center}
\end{table}

\begin{figure*}
\centering
\includegraphics[width=\textwidth]{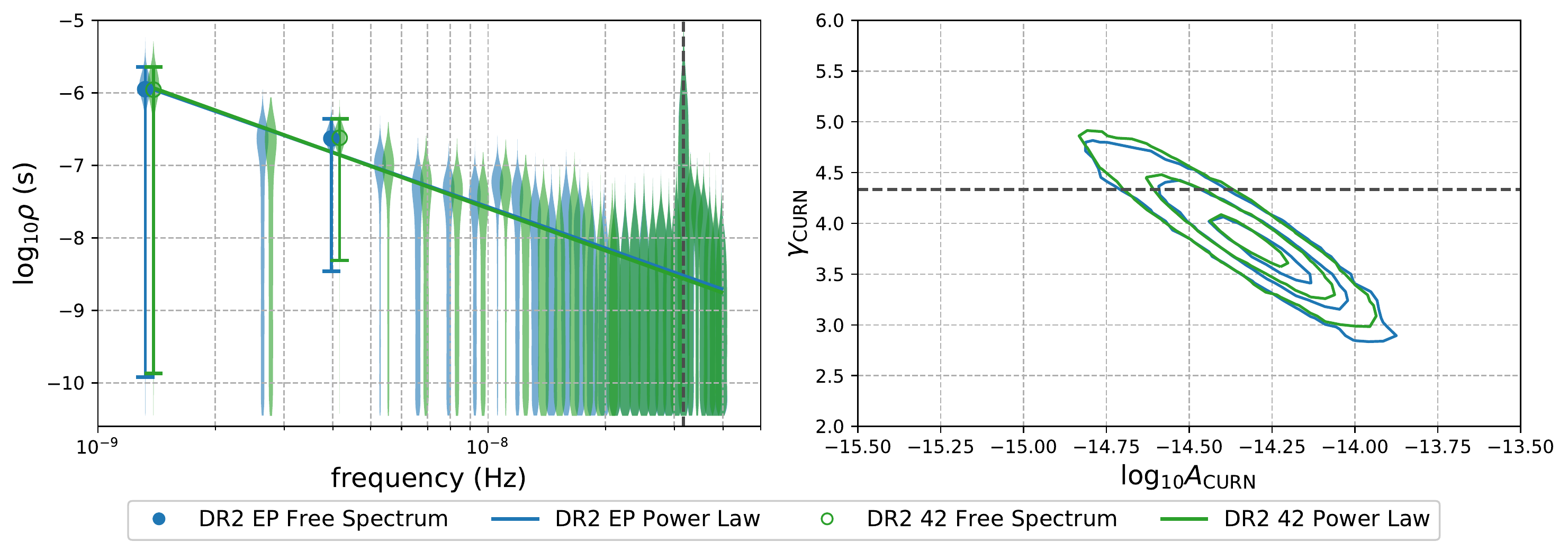}
\caption{Results from CURN analysis using \eprise{} 
(EP) and \ftwo{} (42), both for the free-spectrum (left 
panel) and the single power-law (right panel) analyses. 
The left plot shows the posterior distribution for the 
amplitude of the power at each frequency bin using violin plots. 
Where the inference provides good measurement of the power, 
we denote the median with a circle and the 95\% uncertainties. We consider the measurement good, if more than 95\% of the posterior probability lies above the lowest 6.25\% of the prior.
As the results of the two algorithms are almost identical,
we slightly shift the 42 distributions of the lowest 15 frequency bins
for easier visual comparison.
The right plot shows the two-dimensional posterior 
distribution for the CURN power-law amplitude and spectral index with the dashed line indicating the expected $\gamma=13/3$ from a GWB from SMBHBs. 
The two analysis pipelines have produced consistent results. 
\label{fig:cmp_gw}}
\end{figure*}

\begin{table*}
\caption{95\% constraints on the power-law (PL) parameters for the different analyses discussed in Sections \ref{sec:GWB} and \ref{sec:test_curn} with the Jenson-Shannon divergence computed relative to the \eprise{} fixed SSE run.}
\begin{center}
\def\arraystretch{1.5}
\begin{tabular*}{0.6\textwidth}{c||cc|cc}
\hline
Algorithm + Model & $\log_{10} A_{\rm CRS}$ & J-S div. & $\gamma_{\rm CRS}$ & J-S div. \\
\hline \hline
\eprise{} + DE438 PL           & $-14.29_{-0.33}^{+0.26}$ & $0$ & $3.78_{-0.59}^{+0.69}$ & $0$  \\
\hline
\ftwo{} + DE438 PL             & $-14.33_{-0.31}^{+0.27}$ & $0.00904$ & $3.87_{-0.60}^{+0.67}$ & $0.00942$  \\
\hline
\eprise{} + DE438 CRS PL         & $-14.32_{-0.39}^{+0.31}$ & $0.00833$ & $3.83_{-0.72}^{+0.82}$ & $0.00806$  \\
\hline
\ftwo{} + DE438 CRS PL         & $-14.39_{-0.43}^{+0.33}$ & $0.04521$ & $3.97_{-0.74}^{+0.89}$ & $0.03929$  \\
\hline
\eprise{} + \beph{} PL         & $-14.32_{-0.37}^{+0.30}$ & $0.00604$ & $3.70_{-0.80}^{+0.78}$ & $0.01586$  \\
\hline
\eprise{} + \egp{} PL          & $-14.42_{-0.41}^{+0.32}$ & $0.07145$ & $3.91_{-0.83}^{+0.85}$ & $0.02482$  \\
\hline
\ftwo{} + \lmoss{} PL          & $-14.41_{-0.54}^{+0.35}$ & $0.06118$ & $3.91_{-0.87}^{+1.06}$ & $0.03656$  \\
\hline
\eprise{} + DE438 broken PL    & $-14.24_{-0.37}^{+0.31}$ & $0.02293$ & $3.67_{-0.71}^{+0.76}$ & $0.02097$  \\
\hline
\end{tabular*}
\label{tab:pl_js}
\end{center}
\end{table*}

\begin{figure*}
\centering
\includegraphics[width=0.8\textwidth]{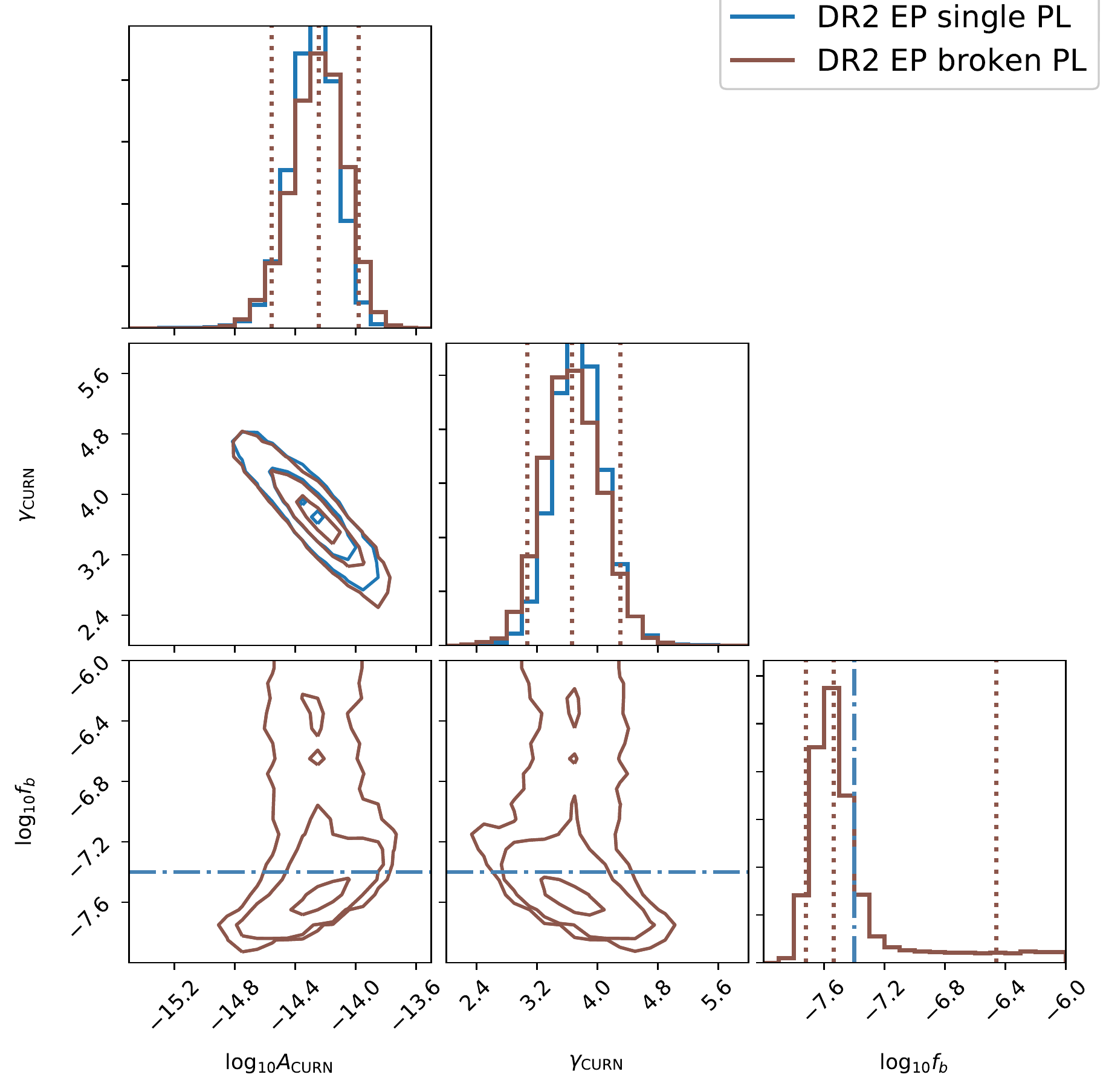}
\caption{Two-dimensional posterior distributions comparing the \eprise{} CURN search with the single and the broken power-law model
with the additional ``bend frequency'' ($f_b$) parameter.
The blue dash-dotted line indicates the
30th frequency bin, 
the highest sampled frequency of the single power-law model. The dotted lines show the median and central 90\% credible regions of the 1D marginalized parameter posteriors. 
The broken power-law analysis
suggests a $f_b$ that corresponds to 20 frequency bins.
The results of the two analyses are completely compatible,
suggesting that the single power-law model with 30 frequency bins describes the data reasonably well.
\label{fig:cmp_fb}}
\end{figure*}

\subsection{Bayesian Model Selection for common red signals}
\label{sec:crs_BF}

The base model to which we make the model comparisons 
is one where the pulsar TOAs only have
independent, uncorrelated
individual pulsar noise (we denote this as PSRN),
without any measurable commonality in the spectral properties
of the different pulsars.
We compare this base model to models that add only one CRS, 
namely either CURN, GWB, CLK and EPH,
as well as models which add two CRSs, i.e. a combination of CURN 
with one of the remaining three ORFs.
The models are listed in Table~\ref{tab:bf_dr2}.
Given the uncertainty on the ORF
this analysis cannot be expected to be fully conclusive,
but can provide indications on whether some of these 
CRS models are more supported by the present data set.

We carried out calculations of Bayes Factors with both the 
\eprise{} and the \ftwo{} packages.
For \eprise{} the Bayes Factors were obtained through a hypermodel structure 
comparing two models against each other.
With \ftwo{} Bayes Factors were calculated using the 
global logarithmic evidence for each analysis,
which is calculated by \pmnest{} 
using the `Importance Nested Sampling' option.
The two analyses give similar results.

Table~\ref{tab:bf_dr2} shows a summary of the Bayes Factors for the different models.
According to the criteria from \cite{kr1995},
the addition of either of the CURN, GWB or EPH signals
to the base PSRN signal
is decisively favoured  
with a $\log_{10}$ Bayes Factor ($\log_{10} {\rm  BF}$) $>2$.
The strongest Bayes Factor is for the CURN model, 
although the evidence for the GWB is only lower by $\log_{10} {\rm  BF}\approx0.4$.
This difference provides only a marginal advantage to the CURN, 
barely disfavouring the GWB signal.
The EPH model, however, is clearly less favoured 
with an $\log_{10} {\rm  BF}$ difference to CURN or the GWB 
of order $\sim1$, which is a substantial difference. 
We will examine the case of EPH in more detail in Section \ref{sec:sseEr_modelselect}.
In contrast to the three models discussed above, 
the monopolar correlation is only mildly favoured
with respect to the PSRN base model.

Since the CURN model has the strongest evidence of the models with a single CRS, 
we can compare it against models which include another 
additional common process. The idea is to test whether there may be 
evidence for several physically motivated common processes 
coexisting in the data. 
In general, none of the three spatially correlated processes add 
substantial evidence to the single CURN.
The ability to distinguish between different spatial correlations could be improved by using more than 6 pulsars in the analysis.
We thus plan to expand the analysis to include 
a larger number of MSPs in the future.

The $\log_{10} {\rm  BFs}$ values obtained with \eprise{} and \ftwo{} 
are very similar and show the same trends. 
The difference in the exact values are within the 
estimated uncertainties by \mnest{}, which are
typically of the order $\sim0.2$ at the 1$\sigma$ level,
suggesting consistency between the two results.
\mnest{} calculates the uncertainty on the 
evidence using the relative entropy of the full sequence of
samples \citep[see][]{ski2004,fh2008}, 
a computationally efficient method that does not require
multiple runs to estimate the variance of the calculated evidence.

In the case of \eprise{}, the uncertainties can be estimated from the number of jumps between the different models \citep{cl2015} or by randomly selecting different sections of the MC chain to get a distribution of Bayes Factors \citep{efron1994}. From one hypermodel analysis run the calculated 1$\sigma$ uncertainty is $\sim0.03$ for both methods.
However, the Bayes Factor values are observed 
to fluctuate at levels about five times higher than the formal uncertainties when running the same analysis multiple times.
Therefore, further investigations are warranted.

\section{Detailed analysis of the common-uncorrelated red noise}
\label{sec:test_curn}

Since the common uncorrelated red noise (CURN) is the favourable model without sufficient
evidence to justify the additional inclusion of another physically
motivated CRS such as a GWB, we proceed with investigating the CURN
more closely. In principle, the same tests can be performed in the 
case where we would be dealing with a GWB. 
In this section, we will specifically examine:
(i) the frequency spectrum of the signal, 
(ii) the effect of the choice of frequency bins when modelling the signal with power-law spectra,
(iii) the stationarity of the signal and consistency between DR1 and DR2 inferences,
(iv) the consistency of individual pulsar noise with the common signal, and
(v) the possible effects from SSE inaccuracies in the CRS analysis.

The power-law spectral parameters inferred 
from all the different CURN-related analyses can be found in Table \ref{tab:pl_js},
along with Jenson-Shannon divergence calculations that compare the results.

\subsection{Individual-frequency modelling Vs Power-law spectrum}
\label{curn_ffr_results}

In order to further investigate our results, we use the 
case of the CURN to investigate the CRS spectrum modelling.
We therefore proceed to also perform the analysis with
an alternative approach to the power-law spectrum model, where the power of each individual CRS-spectrum 
frequency bin is sampled independently.
This approach has been employed in LTM15 and \cite{abb+2020},
and was first discussed in \cite{lah+2013}. 
We refer to this as the `free spectrum` analysis.
We conducted the analysis employing both \eprise{} and \ftwo{},
which provided fully consistent results.
We note that for the \ftwo{} analyses, we used the  \pmnest{} sampler
when implementing the power-law spectrum model and
\pchordlite{} for free-spectrum analyses due to the problem's high dimensionality.
The full posterior distributions can be found in the appendix \ref{fig:cmp_psr}. A comparison with the single-pulsar noise analysis from Table \ref{tab:spna_dr2_params} shows the absorbtion of the pulsar red noises into the CURN, while the DM noises remain relatively consistent.
Figure \ref{fig:cmp_gw} shows the main CURN results of these analyses.
The left panel of Figure \ref{fig:cmp_gw} shows the power of the CURN at each frequency, the free spectrum, with the straight lines indicating the median values of $A$ and $\gamma$ of the posterior distributions from 
the power-law spectral analyses with \eprise{} and \ftwo{} respectively. 
The full 2D posterior contours for 
the power-law parameters are shown on the right panel.
Having confirmed the agreement of the two algorithms, 
we will be using \eprise{} in the rest of the work in this section, except Section \ref{sec:sseEr}.

The free spectrum figure in general has two features. 
At high frequencies
the power is white-noise dominated and can thus be modelled with a flat horizontal line. 
The presence of red noise becomes obvious at the lowest frequency bins and appears to be dominant for about 10 frequency bins.

The posterior distributions on the parameters of the power-law model of the CRS are found to be $\log_{10} A = -14.29_{-0.33}^{+0.26}$ and $\gamma = 3.78_{-0.59}^{+0.69}$ (95\% credible regions), as seen in the right panel of Figure~\ref{fig:cmp_gw} and Table \ref{tab:pl_js}.

\subsection{Choice of the number of power-law frequency bins}
\label{sec:curn_broken}

The simple power-law model can be modified to include a smooth transition from red to white noise. 
This allows us to use the data to determine how many frequency bins are 
needed to optimally sample the low frequency common red noise and which higher 
frequencies are likely to be white noise dominated. We can replace equation~\eqref{eq:gwb_S} with a broken power-law \citep{abb+2020},
\begin{equation}
    S_{\rm CRS} = \frac{A_{\rm CRS}^2}{12\pi^2} \left( \frac{f}{f_c} \right)^{\gamma_{\rm CRS}} \left( 1 + \left(\frac{f}{f_b} \right)^{1/\kappa} \right)^{\kappa(\gamma_{\rm CRS} - \delta_{\rm CRS})}\,,
\end{equation}
where $\gamma_{\rm CRS}$ and $\delta_{\rm CRS}$ are the spectral indices in the low and high frequency regimes respectively, $f_b$ is the bend frequency, where the common signal transitions from the red-noise to the white-noise dominated regime.
The transition smoothness is determined by $\kappa$,
which has been fixed at 0.1 for this analysis, but could be sampled over. In order to test which frequencies contribute to the red noise we need to probe the high frequency regime beyond the 30th frequency bin. Thus, a log-uniform prior for $\log_{10} f_b \in [-9,-6]$ is set.

We find that the most likely bend frequency is around the 20th frequency bin $20/T \approx 2.5\times10^{-8} \textrm{Hz}$, 
below which most of the power is concentrated. To be more conservative we opted to use the conventional 30 frequencies for single power-law models,
knowing that we are accounting for the majority of significant frequency bins. Additionally, we have verified that the power-law spectrum parameters are consistent between the single power-law and the low-frequency end of the broken power-law model, as can be seen in Fig.~\ref{fig:cmp_fb}.

\begin{figure}
\centering
\includegraphics[width=0.4\textwidth]{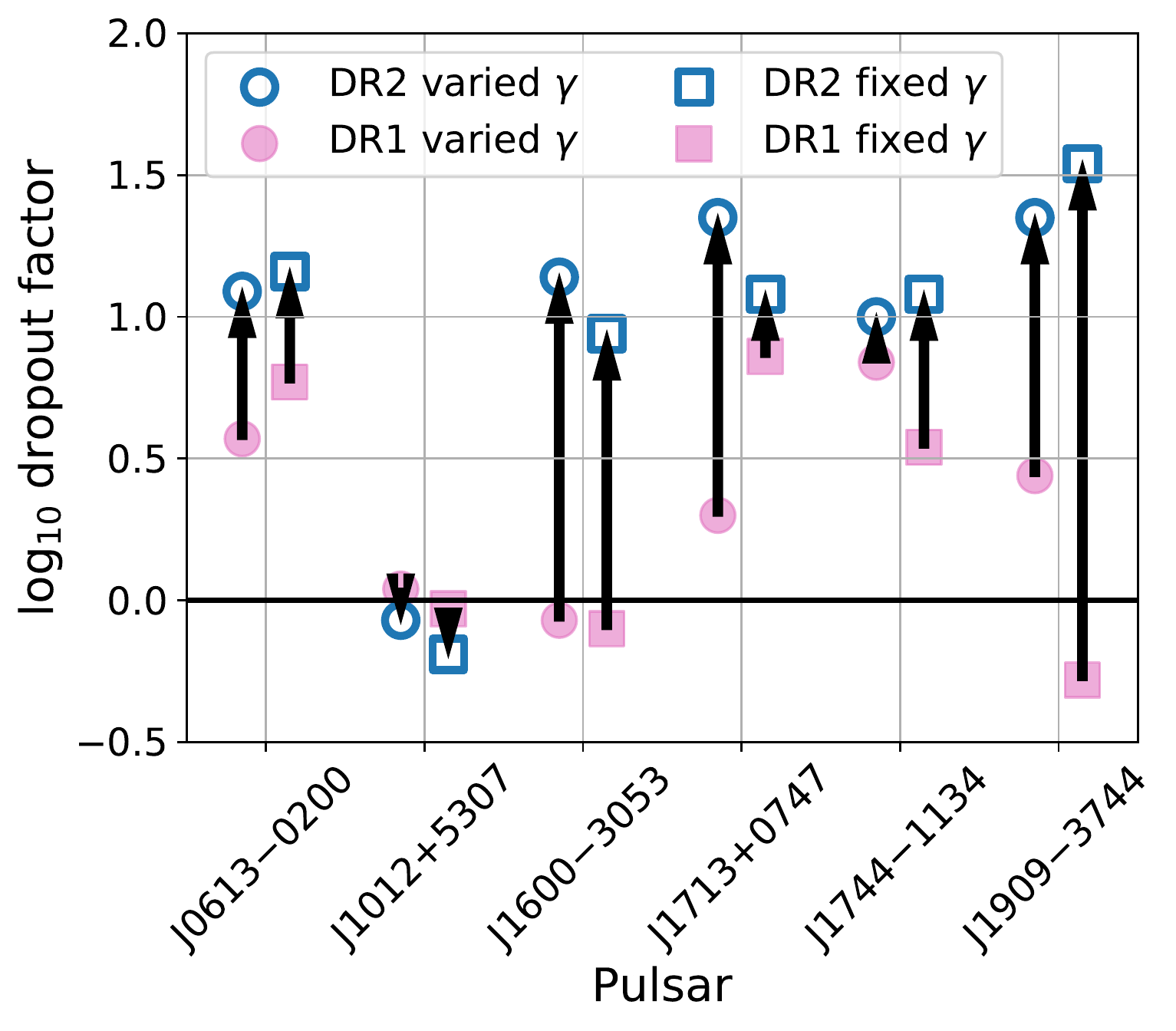}
\caption{Dropout factors for both DR1 and DR2 with varied and fixed $\gamma_{\textrm{CURN}}=13/3$ spectral index shown with circles and squares respectively.
The number of pulsars contributing the CURN detection 
has increased from three to five with the DR2 data extension.
Only PSR~J1012+5307 seems indifferent to the CURN (see Section \ref{sec:disc_compare} for a discussion).}
\label{fig:dropout}
\end{figure}

\begin{figure*}
\centering
\includegraphics[width=\textwidth]{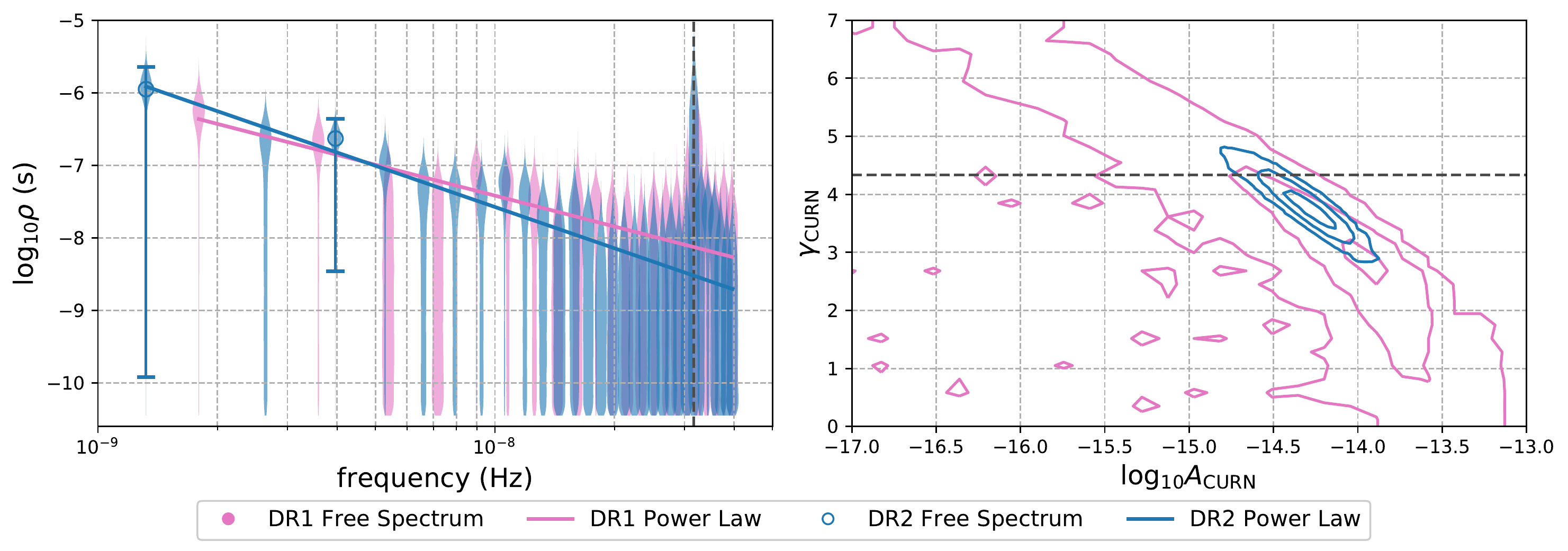}
\caption{Comparison of the CURN recovered signals 
with DR2 and DR1 in the same style as Figure \ref{fig:cmp_gw}, both using free-spectrum and power-law
analyses with \eprise{}. The CURN signal properties
are in agreement with the expected detection evolution
of a stationary red signal when extending the timespan.
The improvement is largely due to the significant increase
in data quality with the DR2 extension.}
\label{fig:fs_dr1dr2}
\end{figure*}

\subsection{Consistency between individual pulsars and the common signal}
\label{sec:removeApsr}

In order to find out how much each pulsar's red noise is consistent with the CURN or in other terms how much a given pulsar contributes to the CURN we employ the 'dropout' method \citep[e.g.][]{aab+2019,abb+2020}.
The dropout factor is defined as,
\begin{equation}
    \begin{split}
    \textrm{dropout}_{k} = & \frac{p_k(\textrm{CURN})}{p_k(\textrm{no CURN)}} \times \\
    & \int \frac{p(\theta_{\textrm{CURN}} | d_{\textrm{no } k}) p(\theta_{\textrm{CURN}} | d_k)}{p(\theta_{\textrm{CURN}})} d\theta_{\textrm{CURN}}\,,
    \end{split}
\end{equation}
where $\theta_{\textrm{CURN}}=(A_{\textrm{CURN}}, \gamma_{\textrm{CURN}})$ is a vector denoting the amplitude and spectral index of the CURN,
$p_k(\textrm{CURN})$ and $p_k(\textrm{no CURN)}$ 
are the probabilities that pulsar $k$ supports or rejects the CURN. 
The terms $p(\theta_{\textrm{CURN}} | d_{\textrm{no } k})$, $p(\theta_{\textrm{CURN}} | d_k)$ 
and $p(\theta_{\textrm{CURN}})$ are the probabilities 
of a CURN with certain spectral properties without pulsar $k$, from pulsar $k$ alone and the overall probability respectively.
The spectral index $\gamma_{\textrm{CURN}}$ can be fixed to $13/3$ for simplicity\footnote{This also allows for an easier decoupling between the pulsar intrinsic red noise and the CURN.}. The integral then becomes a function of only the amplitude $A_{\textrm{CURN}}$.
In summary, the dropout factor is a measure of how much a given 
pulsar $k$ supports the existence of the common signal. It can also be seen as a consistency factor comparing the intrinsic red noise of pulsar $k$ against the CURN constrained by the other pulsars. 
We can use the Bayesian framework from section \ref{sec:crs_BF} to estimate the dropout factors by computing how much each pulsar favours the CURN in a model comparison with the full data set.

The values of the dropout factor for each pulsar for a CURN with freely varying parameters (circles) and fixed $\gamma_{\rm CURN}=13/3$ (squares) can be seen in Figure~\ref{fig:dropout}. A dropout factor of about 1 (or 0 in the logarithmic scale, as in Figure~\ref{fig:dropout}) indicates that the pulsar is indifferent to the CURN. A large dropout factor indicates strong commonality between the pulsar and the CURN. Five out of the six pulsars are in 'support' of the CURN in DR2. PSR~J1012+5307 is noticeably not and will be further discussed in Section \ref{sec:disc_compare}.

Differences between the dropout factors for the same pulsar and data set depending on whether $\gamma_{\textrm{CURN}}$ is fixed or not can be seen in Figure~\ref{fig:dropout}. As the dropout factor can be viewed as a consistency factor, we can compare the intrinsic red noise of a pulsar against the overall constraints on the CURN. E.g. from Table~\ref{tab:spna_dr2_params} we find the median red noise $\gamma_{\rm RN} \sim 3.4$ for PSR~J1713+0747, while the overall CURN has $\gamma_{\rm CURN} \sim 3.8$ (see Figure~\ref{fig:cmp_gw}).
One can expect the single pulsar red noise of PSR~J1713+0747 to be slightly more consistent with the varied CURN posterior than a distribution fixed at $13/3$, thus giving a slightly larger dropout factor when varying $\gamma_{\rm CURN}$.
These differences become more pronounced using DR1, as the constraints on the CURN are tighter in DR2, such that the slice at $\gamma_{\rm CURN}=13/3$ is more representative of the recovered 2D CURN posterior with DR2 in contrast to DR1.

\subsection{Consistency with DR1}

In this work we have added a substantial amount of 
more precise data
to the DR1 data. Therefore we 
investigate whether the CURN properties are consistent between DR1 and this new, extended data set.
If the CURN is stationary,
the analysis of the two data sets ought to produce consistent results, where we should get better constraints with the added data.
This is indeed an important test in the framework 
of searching for a stochastic GWB,
as the signal is theoretically expected to be stationary. 
As DR1 is a very well studied data set, 
it is straightforward to confidently make this investigation. 
We repeated the single-pulsar analysis for DR1 as with DR2,
using the same SSE (DE438) and terrestrial time-standard (BIPM2019)
in order to have an appropriate comparison.
We performed the DR1 CURN power-law analysis using
22 frquency bins, as this was found to be adequate in LTM15.
Again, we used both \eprise{} and \tn{} 
for the single-pulsar noise analysis
and cross-checked the CURN analysis with \eprise{} and \ftwo{}.
As the result with both codes are compatible,
we use the \eprise{} results here to make the
comparisons of the DR2 and the DR1 subset.

Figure~\ref{fig:fs_dr1dr2} shows how the common signal 
has evolved from EPTA DR1 to DR2,
using the posterior distributions of the single power-laws and free-spectra
parameters. One can see that DR2 provides a much more constrained probability distribution
of the power-law parameters. 
While the DR1 data set shows a CURN centered around $\gamma = 2.83_{-1.96}^{+2.14}$ and $\log_{10} A = -13.96_{-1.41}^{+0.34}$ (95\% credible region), 
there is considerable uncertainty in the parameter space beyond
the 95\% credible region.
The additional data from the DR2 data set 
constrain the spectral index closer to the expected value of $\gamma=13/3$ from a GWB by SMBHBs. The amplitude has decreased, also more in line
with more probable theoretical expectations \citep[e.g.][]{csc2019,msc+2021}.
The DR2 free spectrum on the left of Figure \ref{fig:fs_dr1dr2} also seems to be extending the DR1 free spectrum. 
In DR1, about four of the lowest frequencies support the existence of a CURN. 
The median DR2 power-law also passes through the DR1 free spectrum power distributions.

While the timespan extension has contributed to 
the improvement of the CURN analysis, 
we note that this also appears to be to a large degree 
the result of the much better multi-frequency coverage of the newly added 
data. This resulted in very significantly improved 
constraints of the pulsars' DM parameter spaces
and decorrelation of said DM parameters 
from the pulsar red noise parameters.
This is in contrast to DR1, 
where the DM and red noise parameters were significantly correlated
for multiple pulsars, adding uncertainty to the pulsar
red noise parameters that would subsequently result 
in similar uncertainties of common red signals.
We can see how much pulsars have improved in their ability to contribute 
to the recovered CURN, by examining the changes in the dropout factors for each pulsar,
as presented in Figure~\ref{fig:dropout}. 
PSR~J1909$-$3744 is the most prominent
example of the achieved improvement, as it has moved from having the smallest 
contribution to the largest. 
This pulsar has the highest TOA precision, however
in DR1 it only had a time-span of 9.38~yr (in contrast to 15.7~yr in DR2)
and had highly correlated red and DM noise parameters.
The decorrelation of red and DM noise components is achieved thanks
to the wide bandwidth of NUPPI 
(as mentioned in Section~\ref{sec:data}, for this MSP we only use NRT data).
Four other MSPs have increased their dropout factors, supporting the stationarity assumption of the CURN.

We finally examine if the extension of the data set from 
DR1 to DR2 creates any unexpected differences in the Bayes Factors
between the different models examined in Section~\ref{sec:modelselect}.
For the CURN case, and using the DE438 SSE, 
the $\log_{10} {\rm  BF}$ has increased from $\approx1.2$ to $\approx3.7$,
further supporting the stationarity assumption,
and strongly suggesting that the signal, irrespective of its
origin and interpretation, is not a 
statistical fluctuation. We finally note
that despite increased Bayes Factors for the different CRS
signals in DR2 by comparison to DR1, the difference in the 
evidence between CURN and the GWB,
has not drastically change from DR1 (see LTM15), 
thus still not allowing to 
support the finding of a GWB or other spatially correlated signal.
This is most likely due to only using six pulsars in both cases,
which does not offer the necessary sampling of the angular separations.
We also note that the clock-error signal remains the least
favourable physically motivated CRS. 
This is expected from the posterior distribution
of the ORF in Fig.~\ref{fig:gen_crs_dr2}, which is
consistently away from 1 across the pulsar angular separations axis.
The full comparison of Bayes Factors between 
DR1 and DR2 can be found in Table~\ref{tab:bf_dr1_dr2}.

\subsection{Addressing possible common red signals from Solar-system ephemeris systematics}
\label{sec:sseEr}

Previous studies \citep[e.g.][]{thk+2016,gll+2019,vts+2020} 
have shown that the SSE modelling plays an important role 
in the search for common signals with PTA data. 
We therefore investigate the degree by which SSE inaccuracies 
affect the CURN parameter estimation, 
and whether modelling possible SSE-induced signals
affects the CRS model selection results.
In this study we apply three 
independently developed algorithms 
that introduce modelling of the SSE uncertainties 
into the CRS search. This lays the groundwork 
for a robust and cross-checked mitigation of the SSE effects
in future GWB searches.
All three algorithms assume that the SSE parameters 
are close to the correct ones and as such investigate linear deviations
from their values. The algorithms differ in the method used to
derive the induced TOA delays by SSE parameter inaccuracies and the
SSE used as reference. 

The first method applies the \beph{} model \citep{vts+2020}, 
which has previously been used in studies 
estimating upper limits for the GWB and examining common signals \citep{abb+2018,abb+2020}.
This algorithm is based on a physical model that accounts for induced TOA delays
due to linear deviations in planetary masses, 
rotation rate about the ecliptic
pole,
and planetary average orbital elements, 
resulting in a quasi-Keplerian model for the orbit.
Allowing these parameters to vary 
with reference to the SSE DE436 \citep{fp2018}
create what we refer to as variational partials.
The linear combination of partials 
that minimize the differences in the orbit
with respect to DE436 
(differences well below detectability by TOA precision of real data) 
define the values for the planetary parameters
that are used as the \beph{} initial values.
In this work, \beph{} includes terms accounting for the masses of Jupiter, Saturn, Uranus and Neptune, 
the rotation rate around the ecliptic pole and orbital elements for Jupiter as well as for Saturn,
since, the EPTA DR2 is approaching 25 years of timespan.
Each of these terms is linearly perturbed around 
the initial values adding a delay to the TOAs.
The overall linear delay 
(calculated by projecting the partials on the TOAs) 
is treated as a deterministic signal.
The SSE model parameters are MC sampled together with the CRS and pulsar-noise parameters in the Bayesian framework.

In addition to \beph{}, in this work we also
use two other algorithms to control for the effects of SSE errors,
namely \egp{} and \lmoss{}.
For both, this is the first time they have been used 
in the context of any CRS analysis
and this work provides a first comparison 
of their performances against \beph{} with real data. 
While \egp{} is similar to \beph{}, it implements a different model from independent SSE information. \lmoss{}, however, is different from both models,
as it has access to the full equations 
and input parameters of a published dynamical SSE model.

\egp{} (Chalumeau et al., in prep.), describes SSE uncertainties as a Gaussian process. 
The SSE design matrix is based on the partials derived 
using the INPOP19a SSE fit \citep{fdv+2019}, which are mapped into the TOAs via the induced delays.
In this work, we use \egp{} paired with the \eprise{} package and fit for 
the parameters of Jupiter and Saturn orbital elements in the Bayesian search of the CURN. The modified residuals correspond to the deviations in the initial orbital elements of Jupiter and Saturn and mimic the time varying uncertainties in the position of the Solar-system barycentre. Similarly to \beph{}, we have sampled \egp{} parameters together with CURN and pulsar noise parameters. The addition of this analysis allowed for a cross-check of the \beph{} results, within the same analysis framework of likelihood estimation and MC sampler.

\lmoss{} \citep{gll+2019} is integrated with \ftwo{},
and is a fully dynamical model of the major Solar-system bodies
based on the PMOE SSE \citep{ln2003,lzx+2008}. 
This ephemeris was used to optimize the orbit \citep{lyh+2008}
of the space-based GW observatory ``Laser Interferometer Space Antenna'' \citep[LISA;][]{aab+2017}.
LINIMOSS is built making modification in PMOE to match the initial conditions used in the DE435 SSE.
In \cite{gll+2019} it has been 
demonstrated that LINIMOSS/PMOE is compatible with
the DE435 SSE, with differences well below our data precision, 
and can therefore be confidently used
for pulsar timing.
The design matrix for the planetary parameters
is directly derived by first linearly perturbing the 
planetary masses and orbital elements
and numerically re-integrating the SSE;
TOAs predicted using the original SSE are then fitted 
with the modified SSE, and
these SSE-induced TOA delays are added as deterministic signals
in the pulsar design matrices.
In this study we use \lmoss{} 
to analytically marginalize the SSE mass 
and orbital parameters for Jupiter and Saturn 
together with the rest of the timing models during
the search for common signals.

\subsubsection{Effects on the parameter estimation of the common signal}
\label{sec:sseEr_params}

We first focus on the CURN parameter estimation and produce their posterior distributions with the three different algorithms.
The analyses with \beph{} and \egp{} 
perform MC sampling on the SSE parameters
and priors are therefore carefully defined.
Both methods determine the prior range phenomenologically 
by allowing the parameters 
to vary enough to cover differences between various SSE models, 
as well as keeping 
the resulting residuals of the pulsar TOAs 
below a certain threshold to stay within 
the linear regime. For \beph{} the delays are limited to about $\mu$s level, while \egp{} allows for delays from SSE systematics up to about 100 $\mu$s.
Both use uniform priors for the orbital elements, \beph{} uses Gaussian priors for the planetary masses, while they are held fixed in this analysis with \egp{}, after confirming that 
no mass-error signals could be detected by the pulsar data.

\lmoss{} analysis performs analytical marginalization
of planetary masses and orbital elements together with the pulsar timing model
using uniform infinite priors,
making this the analysis with the wider priors.
We note that this analytical marginalization of the SSE parameters 
used here is not the only way it is possible to use \lmoss{}, 
as in principle we may marginalize with specified prior types and ranges, 
or also Monte-Carlo sample these parameters together with the CRS parameters
during the Bayesian inference process.
In this study, we use the full analytical marginalization
as a complementary analysis to the MC sampling used by
\beph{} and \egp{},
and will present the details of general LINIMOSS use in GWB searches
separately (Guo et al., in prep.). 
This has certain limitations in the model selection
process, but also serves as a useful check to see whether
our analyses produce the expected results,
as we discuss below in Sections~\ref{sec:sseEr_params}~and~\ref{sec:sseEr_modelselect}.

\begin{figure}
\centering
\includegraphics[width=0.9\linewidth]{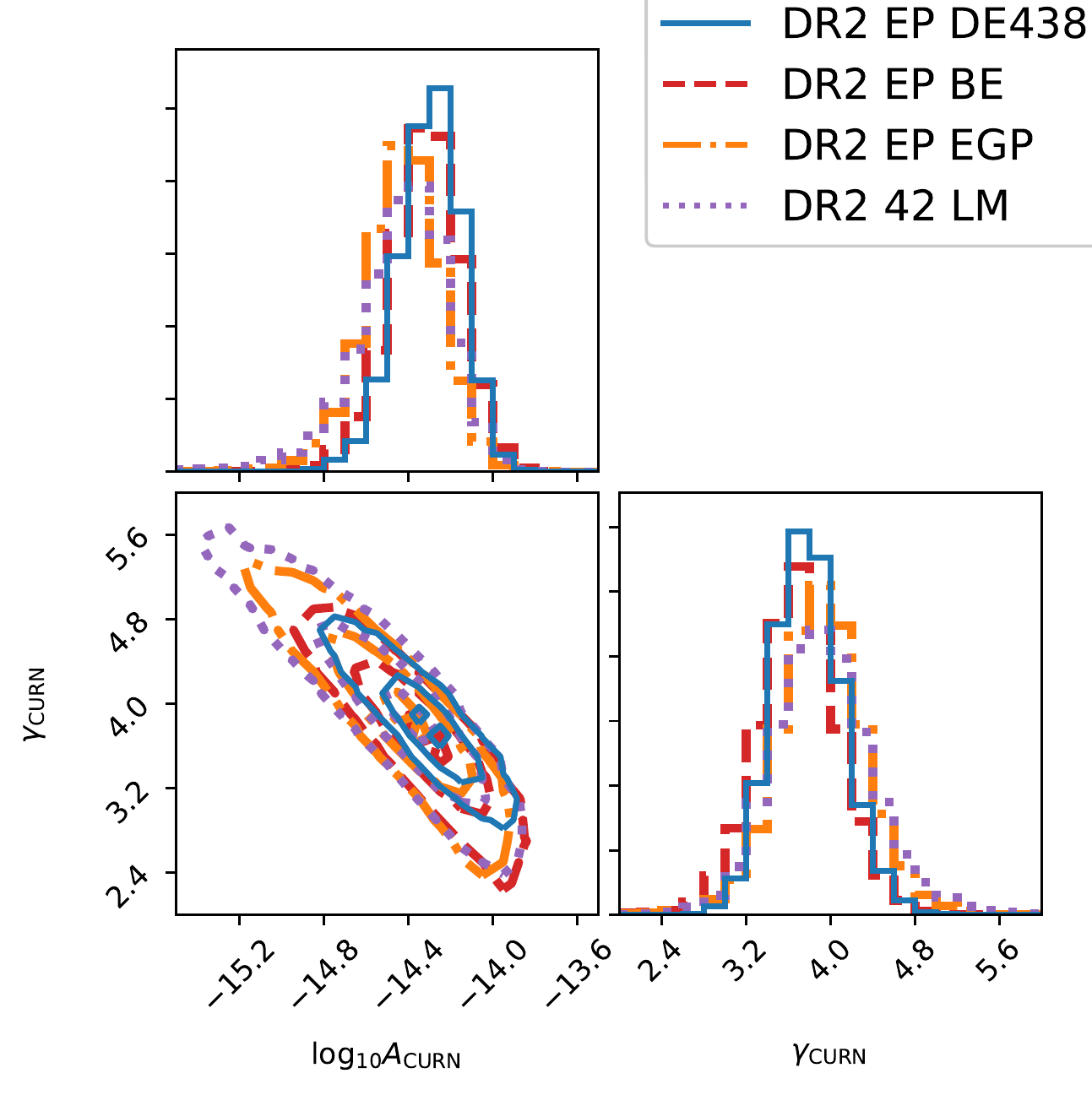}
\caption{Posterior distributions of CURN power-law
parameters using \eprise{}
with (i) fixed DE438 SSE, (ii) \beph{} and (iii) \egp{} and \ftwo{} with \lmoss{}. \beph{} and \egp{}
sample the mass and orbital parameters of Jupiter and Saturn using partials derived from DE436 and INPOP19a. The \lmoss{} anaysis analytically marginalizes these parameters 
together with the pulsar timing models in a fully dynamical analysis with the PMOE SSE.}
\label{fig:dr2_corner_sse}
\end{figure}

Figure~\ref{fig:dr2_corner_sse} shows the comparison of the \eprise{}
results without any SSE fitting, and with the use of \beph{}, \egp{} and \lmoss{} (while the \lmoss{} analysis uses \ftwo{}, 
note that the fixed DE438 distributions from both \eprise{} 
and \ftwo{} are nearly identical, see Fig.~\ref{fig:cmp_gw}, 
and therefore it is sufficient to only show one fixed DE438 result).
We can see that all three methods show consistent 
posterior distributions. 
The inclusion of the 
SSE models slightly increases 
the uncertainties in the recovered parameters, while 
still keeping them highly confined. 
As expected, the contours become progressively broader than the DE438 contour, 
as the allowed prior increases from \beph{} to \egp{}, 
to the full marginalization in \lmoss{}.

\subsubsection{Effects on the model selection for common red signals}
\label{sec:sseEr_modelselect}

The next step is to examine how
including the SSE parameters in the analysis affects the results for the model
selection, as discussed in Section~\ref{sec:modelselect}.
We first examine this in the framework of \eprise{}+\beph{}. The \beph{} model acts as an additional common signal to all pulsars in the array with a dipolar nature. 
As such, it is possible to do a Bayesian model comparison between a CURN analysis (or indeed any of the models listed in Table~\ref{tab:bf_dr2}) with a fixed SSE and one using \beph{}. 
We find only a small $\log_{10} {\rm  BF} \sim 0.4$ in favour of the addition of \beph{} to the PSRN model, while for the CURN model the addition of \beph{} is disfavoured by approximately the same number. Although the addition of \beph{} models the data better than pulsar noise can alone, a simpler CURN provides an equivalently good fit to the TOAs and is therefore the more preferred model.
The small $\log_{10} {\rm BFs}$ indicate that the TOAs are not strongly dependent on any (possible) SSE-parameter inaccuracies.

\begin{table}
\caption{Results from model selection analysis, in logarithmic (base 10) Bayes Factors ($\log_{10} {\rm  BF}$), for different CRS models, using \eprise{} and fitting SSE parameters with \beph{}. The model IDs and acronyms are the same as in Table~\ref{tab:bf_dr2}. PSRN has no $\log_{10} {\rm  BF}$ values as it serves as the base model.}
\begin{center}
\def\arraystretch{1.5}
\begin{tabular*}{0.7\linewidth}{c|c|c}
\hline
ID & Model & \eprise{}+\beph{} \\
\hline \hline
0 & PSRN                 & -- \\
\hline
1 & PSRN + CURN          & $2.9$ \\
\hline
2 & PSRN + GWB           & $2.7$ \\
\hline
3 & PSRN + CLK           & $2.3$ \\
\hline
4 & PSRN + EPH           & $1.0$ \\
\hline
\end{tabular*}
\label{tab:bf_sse_be}
\end{center}
\end{table}

\begin{table}
\caption{Results from model selection analysis, in logarithmic (base 10) Bayes Factors ($\log_{10} {\rm  BF}$), for different CRS models, using \ftwo{} and fitting SSE parameters with \lmoss{}.The model IDs and acronyms conventions are the same as in Table~\ref{tab:bf_dr2}. PSRN has no $\log_{10} {\rm  BF}$ values as it serves as the base model. Note that these results are not directly comparable to 
those in Table~\ref{tab:bf_sse_be} as explained in the main text.}
\begin{center}
\def\arraystretch{1.5}
\begin{tabular*}{0.7\linewidth}{c|c|c}
\hline
ID & Model & \ftwo{}+\lmoss{} \\
\hline \hline
0 & PSRN                 & -- \\
\hline
1 & PSRN + CURN          & $1.3$ \\
\hline
2 & PSRN + GWB           & $1.3$ \\
\hline
3 & PSRN + CLK           & $1.2$ \\
\hline
4 & PSRN + EPH           & $0.0$ \\
\hline
\end{tabular*}
\label{tab:bf_sse_lmoss}
\end{center}
\end{table}

The $\log_{10} {\rm  BFs}$ when including \beph{} in the analysis for the same models as in Section \ref{sec:modelselect} can be found Table \ref{tab:bf_sse_be}. 
As we have seen that models with two CRS components in addition to PSRN
are not an improvement to models with only one CRS added,
we only focus on the latter models. The evidence for CURN, GWB remain very significant even if we add uncertainty modelling into the SSE. However, the $\log_{10} {\rm  BF}$ for the EPH CRS is significantly lower. This is an expected result as \beph{} is designed to take care of the possible systematics in the SSE encoded in the Solar-system barycentre. The drop in significance of the (PSRN+EPH) model corresponds to the absorption of dipolar correlations by \beph{}.
At the same time, the CLK model now has a much stronger support.
This suggests that after modelling the SSE parameters 
the ORF may become flatter
and more similar to a monopolar signal, and that indeed the use of only six pulsars
makes it difficult to distinguish between different angular correlation shapes.

In addition to the model selection process using \beph{},
we also perform an analogous analysis using \lmoss{}.
It differs from the \beph{} analysis in that we cannot
discern if a model selection would prefer a fixed SSE or the inclusion of SSE fitting with \lmoss{}.
This is because the analytical marginalization of the SSE parameters 
implies the use of improper (un-normalizable) priors that 
makes the model with and without the SSE fit not directly comparable.
Nevertheless, we may still perform Bayesian model selection, 
if all models in question are affected by the marginalization over the improper prior. 
We therefore additionally compare the same models reported in the Table \ref{tab:bf_sse_be}, but now using \lmoss{} and analytic marginalization
(infinite priors) over the planetary parameters.
Table~\ref{tab:bf_sse_lmoss} overviews these results.
There are two important observations one can immediately make.
First, the model (PSRN+EPH) is significantly disfavoured by comparison to other CRS models as \lmoss{} 
fully absorbs any dipolar correlations related to SSE signal in the data
with the analysical marginalization. 
This is an expected and highly desired result.
The second observation is that the Bayes Factors for the 
CURN and GWB are reduced more
than in the case of \beph{}, and in addition,
the evidences for the CRS to be a CURN, 
a GWB or a CLK signal are equalized, following a similar
trend as with \beph{}, but more prominently.
These differences from the \beph{} 
results are most likely because
the full marginalization of the SSE parameters,
while only using six pulsars in the analysis,
can lead to indiscriminate
absorbtion of other types of CRS, correlated or not.
Tests using simulated data have confirmed this scenario.
The equal evidences in CURN, a GWB or a CLK signals
after fitting the SSE with \lmoss{} 
also supports that the data with six pulsars cannot distinguish
the nature of any remaining CRS after fitting for the SSE.
A more detailed analysis using \lmoss{} with more precise
techniques in order to fully interpret the results,
will be published separately (Guo et al., in prep.).

We also plan to compute Bayes Factors and perform similar investigations with \egp{} in a future publication (see Chalumeau et al., in prep.). 
As we have access to the full design matrix of INPOP19a, 
more planet masses and orbits can be added to the \egp{} model used in this work. 
The effects of a properly chosen prior range will also be tested.

The general conclusion is that overall the 
SSE analytical marginalization is very good at absorbtion 
of dipolar signals but is less safe with respect to leaving 
a true GWB signal in the data unabsorbed; 
this is in agreement with findings by \cite{thk+2016}.
As such, this type of analysis
is a good basis for a very conservative GWB search
or strain upper limit.
On the other hand, the approach of sampling the SSE parameters
requires careful prior choices as it may leave some dipolar signal
unmodelled, potentially affecting the measurement of parameters
and detection significance of a true GWB signal. 
These issues can be reduced significantly with more pulsars to better
sample the angular separations,
which highly motivates the work for 
the preparation of the upcoming full EPTA DR2 data set.

\section{Discussion of results and comparison with literature}
\label{sec:disc_compare}

\begin{figure*}
\centering
\includegraphics[width=\textwidth]{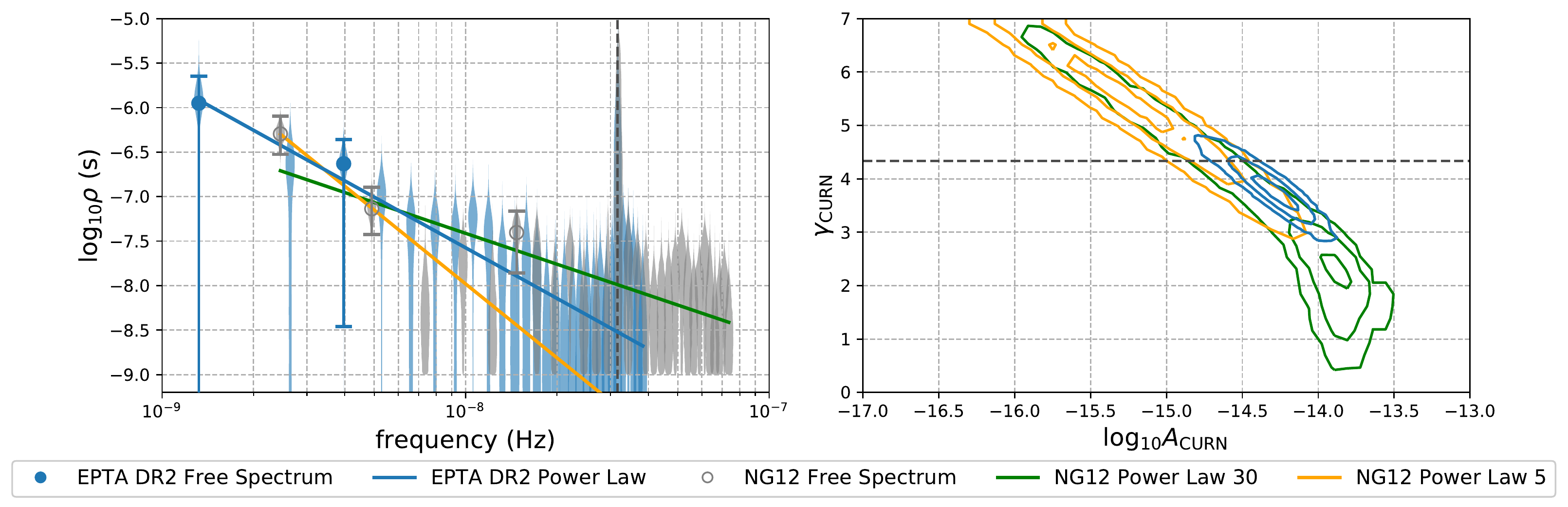}
\caption{Common uncorrelated red noise search free spectrum and power-law recovery comparison between EPTA DR2 and NG12 with DE438 \citep{abb+2020} in the same style as Figure \ref{fig:cmp_gw}}
\label{fig:gwb_dr2_ng}
\end{figure*}

\begin{figure}
\centering
\includegraphics[width=0.4\textwidth]{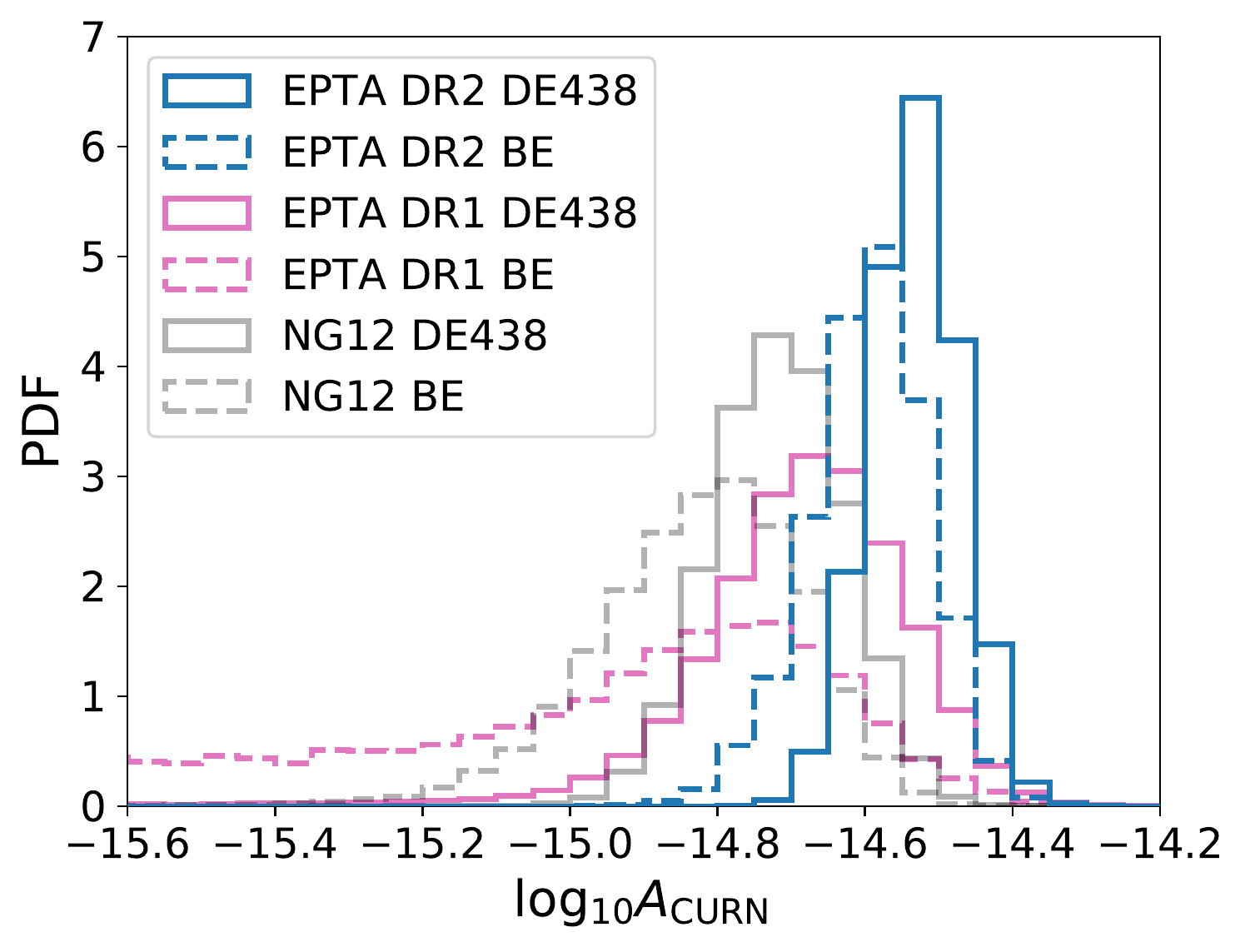}
\caption{Comparison of the CURN recovered signals 
with EPTA DR2, DR1 and NG12 \citep{abb+2020}, using fixed DE438 and \beph{} with \eprise{} and fixed $\gamma_{\rm CURN}=13/3$.}
\label{fig:hist_dr1dr2}
\end{figure}

\begin{figure*}
\centering
\includegraphics[width=\textwidth]{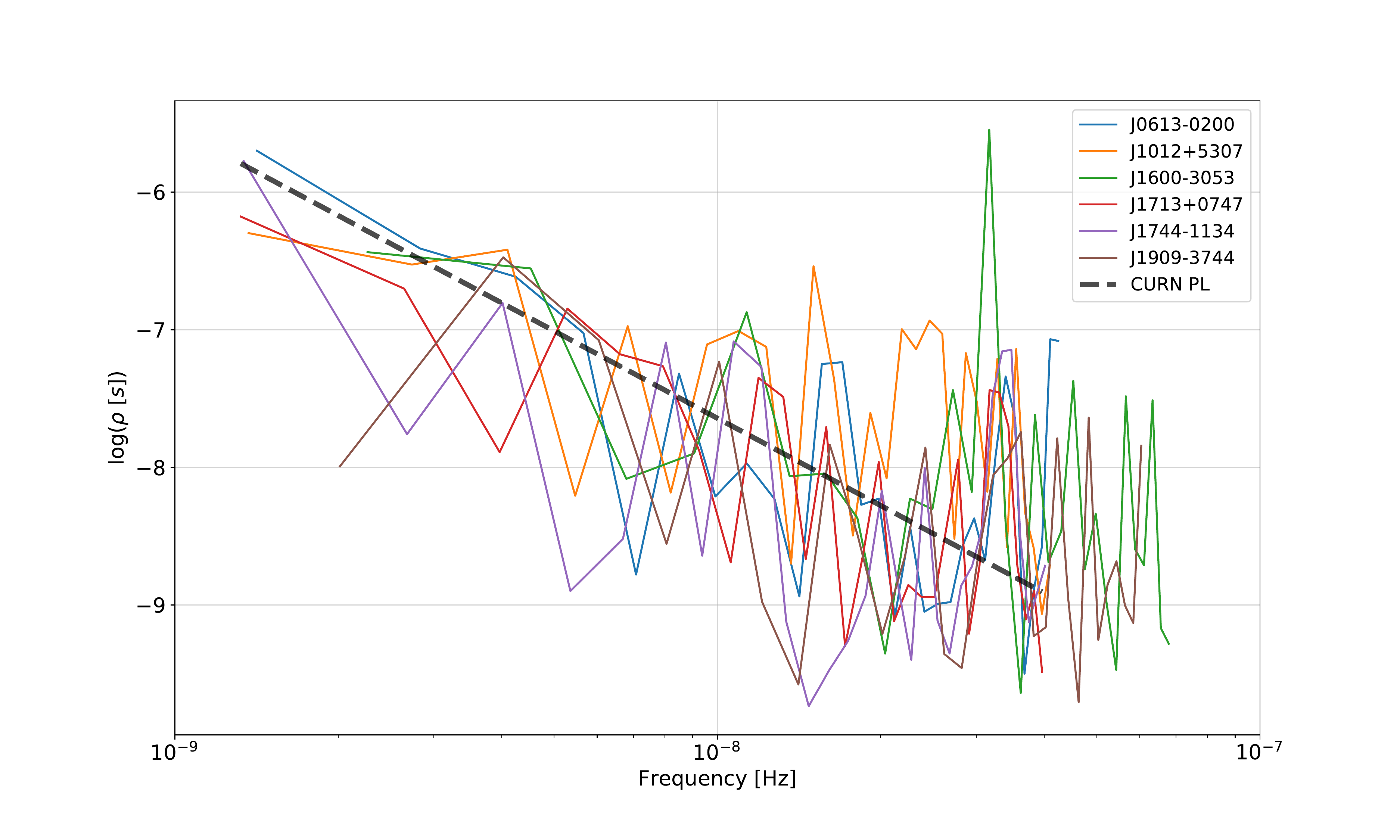}
\caption{Single pulsar noise spectra 
(Maximum aposteriori values from \eprise{} noise analysis) 
for each of the six pulsars, compared to the CURN power-law (median values)
shown with the dashed line.
\label{fig:psr_spectra}}
\end{figure*}

A common uncorrelated red noise (CURN) process has been reported by independent data sets and their analysis by other PTA collaborations \citep[][]{abb+2020,gsr+2021}. 
We will briefly compare the results from this work against the constraints on the power-law parameters of the CURN 
and the model selection Bayes Factors from \cite{abb+2020}. 
A more detailed comparison between the inferences from
the different regional PTA groups will be presented in a relevant upcoming IPTA paper (Antoniadis et al., in prep.).

Fig.~\ref{fig:gwb_dr2_ng} compares the fixed DE438 CURN analysis from the EPTA DR2 against the NANOGrav 12.5yr data set. The left panel shows a broad agreement in the power distributions by frequency. As the EPTA DR2 is almost twice as long as the NANOGrav data, it provides a tighter constraint on the CURN power-law spectral slope, and consequently amplitude. In addition, this result is largely insensitive to the choice of the number of frequency bins used in the analysis, see Section \ref{sec:curn_broken} and Fig~\ref{fig:cmp_fb},
in contrast to NANOGrav's recovered spectral properties.

We can also compare the amplitude of a CURN at fixed $\gamma=13/3$, which corresponds to a GWB from massive black hole binaries. The EPTA DR2 results give an amplitude of
$A_{\rm CURN} = 2.95_{-0.72}^{+0.89} \times 10^{-15}$ (95\% credible region). 
Figure \ref{fig:hist_dr1dr2} shows a comprehensive comparison of the relevant
CURN amplitude posterior distributions when using the DE438 SSE and
when fitting SSE planetary parameters with \beph{}, for EPTA DR2 and DR1 and
the NANOGrav 12.5yr results from \cite{abb+2020}. 
We note that the EPTA results are obtained with an upper prior bound of $\log_{10} A = -10$ and simultaneously fit for the pulsar DM stochastic noise, whereas NANOGrav uses a $\log_{10} A = -14$ upper prior bound and observationally measured piecewise DMX values to model the pulsar DM variation over time \citep[e.g.][]{abb+2015a,jml+2017}.
As such, this comparison can only be indicative.

In this paper, we have compared power-law models for common red signals (CRSs) with 
different overlap reduction functions (ORFs) while retaining the spectral index as a 
free parameter. We obtain a $\log_{10} {\rm  BFs}$ of $\approx 3.7$ for PSRN+CURN vs PSRN (DE438). In order to make a more direct comparison with the NANOGrav model 
selection results, we have repeated our analysis by fixing the spectral index $\gamma=13/3$ and matching the NANOGrav prior on the common signal amplitude. 
In this case we find $\log_{10} {\rm  BF}$ $\gtrsim4$ compared to the NANOGrav result of 4.5 
(DE438) with an estimated uncertainty of $0.9$. 
We can therefore conclude that the EPTA DR2 and NANOGrav 12.5yr results are in general agreement.

Our EPTA DR2 model comparison results appear to be less affected by SSE  uncertainties than the NANOGrav 12.5yr results. A possible reason is that the NANOGrav data set
has a maximum timespan of roughly half a year longer than the Jovian 
orbital period, while the EPTA data covers close to two Jovian orbits
and is a few years short of Saturn's orbit.
As such, the EPTA data may be more effective
in distinguishing signals induced by errors in Jupiter's orbit from pulsar noise.
Another possibility, however, is that the present EPTA data set
does not efficiently recover the dipolar correlations with \beph{}
due to only using six pulsars, with sparse coverage of the 
pulsars angular separations space. The addition of more pulsars in
the EPTA array will provide more information on this.

\
\\
The commonality of the single pulsar red noises, 
which could be interpreted as a CRS, has been investigated by the PPTA \citep{gsr+2021}. 
Though not unexpected, 
the PPTA has used simulated data to unambiguously demonstrate
that individual pulsar noises can be recovered by the analysis
code as a CURN, if the spectral properties are similar.
This is something that must be considered carefully, 
as MSPs possibly may have common underlying mechanisms 
that produce intrinsic stochastic noise \citep[see e.g][]{jon1990,sc2010,ml2014}.
In Fig.~\ref{fig:psr_spectra}, we show the  
power spectra for each of the 6 EPTA pulsars 
using the maximum a posteriori values of the SPNA runs from \eprise{}. One can see a broad agreement of all pulsars with the CURN, thus strengthening this CURN as a common noise floor. However, PSR~J1909$-$3744's red noise power is poorly constrained at the lowest frequency, consequently, plotting a point estimate can give the impression of a dip below the CURN.
Looking at PSR~J1012+5307, the red noise is clearly present in this pulsar and its shape is consistent with other pulsars. However the level of white noise is higher and, as a result, its slope appears to be lower and thus inconsistent with the CURN. This is a possible cause of results of dropout analysis observed for this pulsar in Fig.~\ref{fig:dropout}. 

The noise properties of the six pulsars used in this study are 
analyzed in greater detail and the results will be published separately in an upcoming paper (Chalumeau et al., in prep.).
In a similar fashion that \cite{lsc+2016} has examined the noise properties of the IPTA DR1 and more recently \cite{grs+2021} of the PPTA DR2,
the EPTA is optimizing the pulsar noise models 
via Bayesian model selection.
One important aspect will be on the chromaticity, i.e. radio-frequency dependence, of the noise. Apart from DM, scattering variation can also introduce a significant noise term to the TOAs \citep{msa+2020}.
Other additional noise components 
such as band noise 
(i.e. noise confined in a specific observing-frequency band)
and system noise
(i.e. noise attributed to one 
specific observing system, therefore not
being intrinsic pulsar noise)
will also be investigated. 
Finding system noise 
requires overlapping data
by multiple observing systems in the same frequency band. 
The addition of new-generation-backened data are now allowing
better such investigations than DR1, 
where only some basic investigation
could be applied \citep[see][]{cll+2016} and further 
work had to be completed on the IPTA DR1 \citep{lsc+2016}.
Despite the DR2 improvement, 
The IPTA remains the best framework to identify possible system noise components.

\section{Conclusions}
\label{sec:conclude}

The EPTA has collected and analyzed observations for six MSPs 
using five radio telescopes for a timespan of up to 24 years.
Data are collected both from single-telescope and phased-array (LEAP)
observations, in a wide range of radio frequencies. The DR1 has been published in \cite{dcl+2016} and analyzed to place an uppper limit
on the isotropic stochastic GWB strain amplitude (LTM15); 
the corresponding amplitude
limit at the 95\% credible region 
for circular, GW-driven SMBHBs was $3\times 10^{-15}$. 
Initial analysis with the DR1 on the correlated search had shown the possible presence 
of a CRS amongst the six MSPs. 
In this work we re-analyze the same six MSPs using
the extended DR2 data and find increased and 
strong evidence for the CRS,
the properties of which are now very well measured 
and remain consistent with the DR1 results.

We have determined the pulsar noise properties 
with two separate pipelines with fully consistent results.
A simultaneous search for the spectral properties 
and spatial correlation of a common red signal detects a well-constrained contour for the power-law spectral parameters 
with $\log_{10}A=-14.32_{-0.39}^{+0.31}$
and $\gamma=3.83_{-0.72}^{+0.82}$ (95\% credible regions),
but no conclusive measurement on the spatial correlation.
Thus, we employed Bayesian model selection
to compare different physically motivated spatial correlations
in their ability to fit the data.
In the case where the SSE is fixed to DE438, 
we find strong evidence for a CURN
in addition to intrinsic, individual pulsar red noise
with a $\log_{10} {\rm  BF}$ $\sim 3.7$ for varying spectral index. 
This is a significant increase from the $\log_{10} {\rm  BF}$ $\sim 1$ in DR1.
The model selection for CRSs with different ORFs
shows very little evidence for a monopolar 
and some evidence for a dipolar signal. The Hellings-Downs correlation is slightly disfavoured compared to the CURN, but more favoured than the dipolar correlation. Nonetheless,
we cannot rule out that the CRS is due to SSE systematics with this current results.
We further analyzed the data sets using three independent SSE-error mitigation models: \beph{}, \egp{} and \lmoss{}. 
While we concluded that our results are not significantly
affected by SSE inaccuracies, we have confirmed that 
including SSE modelling reduces the evidence for a dipolar
CRS in the data, and confirmed that the evidence reduction is 
stronger with wider priors on the values
of planetary mass and orbital parameters.
The effect of the SSE fitting in the general CURN search is a slight increase in the confidence intervals of the amplitude and spectral index of the power-law model. Longer timespan, more precise TOAs and more pulsars can help to diminish the effects of SSE systematics in the CURN recovery.

Assuming a single power-law model for the CURN, the recovered parameters constraints from the EPTA DR2 are
$\log_{10} A = -14.29_{-0.33}^{+0.26}$
and $\gamma = 3.78_{-0.59}^{+0.69}$ (95\% credible regions).
This power-law spectrum model uses the 30 lowest
frequency bins. An analysis with a 'broken power-law' spectrum
has indicated the optimal number of frequency bins for the
red part of the spectrum is $\sim20$.
However, the corresponding 2D posterior contours are consistent with those of the single power-law analysis, confirming the robustness
our results to the exact number of frequency bins.

We have measured the contributions of the individual pulsars to the CURN 
using the dropout method. This produces a factor of consistency between the 
red noise of a given pulsar with the CURN. 
We find support for five out of six pulsars contributing to the CURN.
This is an increase from three pulsars in DR1. 
The overall dropout consistency also increases significantly, particularly for J1600$-$3053 and J1909$-$3744, where DR2 adds the most data relative to DR1.

The nominal amplitude $A_{\textrm{GWB}}$ for a power-law at fixed $\gamma = 13/3$ at 1/yr has been found to be $2.95_{-0.72}^{+0.89} \times 10^{-15}$.
While there is a notable difference in the reported amplitude
median values compared to the NG12 amplitude of $1.9 \times 10^{-15}$ and PPTA DR2 amplitude of $2.2 \times 10^{-15}$, an overlap at the 2-3 sigma-level remains; in particular the significantly large NG12 parameter space for the spectral properties of the CURN cautions one in drawing a conclusion
only from median values.
Differences at this point, nevertheless, could have a multitude of reasons. One particular point could be in the covariance and modelling of the pulsar noise, which could leak into the CURN. Further investigations and cross-comparisons between the different data sets will be done in the IPTA framework.

If the full amplitude of $3\times 10^{-15}$ of the EPTA DR2 is due to the GWB, this would necessitate various strong astrophysical effects \citep[see e.g.][]{ses2013,kbh+2017,skh2020}. Following \cite{msc+2021}, the two main observables are the black hole binary mass and the merger time scale. The hypothetical amplitude of $3\times 10^{-15}$ is close to the upper bound of possible values. In order to achieve such a strong GWB emission, black hole binaries have to be very heavy and the overall merger time scale needs to be short. This would place very stringent limits on current massive black hole formation and evolution models. Alternatively, a number of other GW emission mechanisms can produce such an amplitude, such as cosmic strings \citep[see e.g.][]{oms2010,sbs2012}, primordial black holes \citep[see e.g.][]{gri2005,lms+2016}, phase transitions \citep[see e.g.][]{cds2010,klm+2017} and numerous other sources.

This paper is the first of a series introducing the next generation of the EPTA data set DR2.
In this work, we have focussed on the six MSPs that have been analyzed previously with DR1. Despite the strong evidence for a CRS, 
we have no measurement of the characteristic Hellings-Downs 
spatial correlation of the GWB. This is expected given the small number of pulsars used. The EPTA is preparing the expansion data set to 
$\sim25$ pulsars, carefully selected based on their data quality, noise properties and sky positions to maximize our sensitivity to a GWB with HD ORF.
The EPTA is also working on analyzing the expanded DR2 data set
more carefully regarding all aspects of the analysis, 
including, but not limited to, the pulsar timing and noise analysis, chromatic noise mitigation, Solar-system dynamics modelling and more consistency checks on any CRS signal.

\section*{Acknowledgements}

The European Pulsar Timing Array (EPTA) is a collaboration between European and partner institutes,
namely ASTRON (NL), INAF/Osservatorio di Cagliari (IT), Max-Planck-Institut f\"{u}r 
Radioastronomie (GER), Nan\c{c}ay/Paris Observatory (FRA), the University of Manchester 
(UK), the University of Birmingham (UK), the University of East Anglia (UK), the 
University of Bielefeld (GER), the University of Paris (FRA), the University of 
Milan-Bicocca (IT) and Peking University (CHN), with the aim to provide high precision 
pulsar timing to work towards the direct detection of low-frequency gravitational waves. An Advanced Grant of the European Research Council to implement the Large European Array for Pulsars (LEAP) also provides funding. The EPTA is part of the International Pulsar Timing Array (IPTA), we would like to thank our IPTA colleagues for their help with this paper.

Part of this work is based on observations with the 100-m telescope of the Max-Planck-Institut f\"{u}r Radioastronomie (MPIfR) at Effelsberg in Germany. Pulsar research at the Jodrell Bank Centre for Astrophysics and the observations using the Lovell Telescope are supported by a Consolidated Grant (ST/T000414/1) from the UK's Science and Technology Facilities Council. The Nan{\c c}ay radio Observatory is operated by the Paris Observatory, associated to the French Centre National de la Recherche Scientifique (CNRS), and partially supported by the Region Centre in France. We acknowledge financial support from ``Programme National de Cosmologie and Galaxies'' (PNCG), and ``Programme National Hautes Energies'' (PNHE) funded by CNRS/INSU-IN2P3-INP, CEA and CNES, France. We acknowledge financial support from Agence Nationale de la Recherche (ANR-18-CE31-0015), France. The Westerbork Synthesis Radio Telescope is operated by the Netherlands Institute for Radio Astronomy (ASTRON) with support from the Netherlands Foundation for Scientific Research (NWO). The Sardinia Radio Telescope (SRT) is funded by the Department of University and Research (MIUR), the Italian Space Agency (ASI), and the Autonomous Region of Sardinia (RAS) and is operated as National Facility by the National Institute for Astrophysics (INAF).

The work is supported by National SKA program of China 2020SKA0120100, Max-Planck Partner Group, NSFC 11690024, CAS Cultivation Project for FAST Scientific. This work is supported as part of the ``LEGACY'' MPG-CAS collaboration on low-frequency gravitational wave astronomy. AC acknowledges support from the Paris \^{I}le-de-France Region. GS, AS and AS acknowledge financial support provided under the European Union's H2020 ERC Consolidator Grant ``Binary Massive Black Hole Astrophysic'' (B Massive, Grant Agreement: 818691). JA acknowleges support by the Stavros Niarchos Foundation (SNF) and the Hellenic Foundation for Research and Innovation (H.F.R.I.) under the 2nd Call of ``Science and Society'' Action Always strive for excellence -- ``Theodoros Papazoglou'' (Project Number: 01431). GD, RK and MK acknowledge support from European Research Council (ERC) Synergy Grant ``BlackHoleCam'' Grant Agreement Number 610058 and ERC Advanced Grant ``LEAP'' Grant Agreement Number 337062. JWMK is a CITA Postdoctoral Fellow: This work was supported by the Natural Sciences and Engineering Research Council of Canada (NSERC), (funding reference CITA 490888-16). AV acknowledges the support of the Royal Society and Wolfson Foundation. JPWV acknowledges support by the Deutsche Forschungsgemeinschaft (DFG) through the Heisenberg programme (Project No. 433075039).

\section*{Data Availability}

The timing data used in this article shall be shared on reasonable request to the corresponding authors.


\bibpunct{(}{)}{;}{a}{}{,}
\def\aap{A\&A}                
\def\aapr{A\&A~Rev.}          
\def\aaps{A\&AS}              
\def\aj{AJ}                   
\def\ajph{Australian J.~Phys.}
\def\alet{Astro.~Lett.}       
\def\ao{Applied Optics}       
\def\apj{ApJ}                 
\def\apjl{ApJ}                
\def\apjs{ApJS}              
\def\apss{Ap\&SS}             
\def\araa{ARA\&A}             
\def\asr{Av.~Space Res.}     
\def\azh{AZh}                 
\def\baas{BAAS}               
\def\cpc{Comput.~Phys.~Commun.} 
\def\gca{Geochim.~Cosmochim.~Acta} 
\def\iaucirc{IAU Circ.}       
\def\ibvs{IBVS}               
\def\icarus{Icarus}           
\def\jcomph{J.~Comput.~Phys.} 
\def\jcp{J.~Chem.~Phys.}      
\def\jgr{J.~Geophys.~R.}      
\def\jrasc{JRASC}             
\def\met{Meteoritics}         
\def\mmras{MmRAS}             
\def\mnras{MNRAS}             
\def\mps{Meteoritics and Planetary Science} 
\def\nast{New Astron.}        
\def\nat{Nature}              
\def\pasj{PASJ}               
\def\pasp{PASP}               
\def\phr{Phys.~Rev.}          
\def\pdra{Phys.~Rev.~A}       
\def\prb{Phys.~Rev.~B}       
\def\prc{Phys.~Rev.~C}       
\def\prd{Phys.~Rev.~D}       
\def\phrep{Phys.~Rep.}        
\def\phss{Phys.~Stat.~Sol.}        %
\def\procspie{Proc.~SPIE}     
\def\planss{Planet.~Space Sci.}  
\def\qjras{QJRAS}             
\def\rpph{Rep.~Prog.~Phys.}   
\def\rgsp{Rev.~Geophys.~Space Phys.~} 
\def\sal{Soviet Astron.~Lett.}
\def\sci{Science}             
\def\solph{Sol.~Phys.}        
\def\ssr{Space Sci.~Rev.}     
\def\zap{Z.~Astrophys.}       
\def\jasa{J.~Amer.~Stat.~Assoc.} 

\bibliographystyle{mnras}
\bibliography{psrrefs,references} 



\ \\
	$^{1}$Laboratoire de Physique et Chimie de l'Environnement et de l'Espace LPC2E UMR7328, Universit{\'e} d'Orl{\'e}ans, CNRS, 45071 Orl{\'e}ans, France\\
	$^{2}$Station de Radioastronomie de Nan\c{c}ay, Observatoire de Paris, PSL University, CNRS, Universit{\'e} d'Orl{\'e}ans, 18330 Nan\c{c}ay, France\\
	$^{3}$Kavli Institute for Astronomy and Astrophysics, Peking University, Beijing 100871, P.R.China\\
	$^{4}$Max-Planck-Institut f{\"u}r Radioastronomie, Auf dem H{\"u}gel 69, 53121 Bonn, Germany\\
	$^{5}$Universit{\'e} de Paris, CNRS, Astroparticule et Cosmologie, 75013 Paris, France\\
	$^{6}$Dipartimento di Fisica ``G. Occhialini", Universit{\`a} degli Studi di Milano-Bicocca, Piazza della Scienza 3, 20126 Milano, Italy\\
	$^{7}$INFN, Sezione di Milano-Bicocca, Piazza della Scienza 3, 20126 Milano, Italy\\
	$^{8}$National Astronomical Observatories, Chinese Academy of Sciences, Beijing, 100101, P. R. China\\
	$^{9}$Moscow Institute of Physics and Technology, Dolgoprudny, Moscow region, Russia\\
	$^{10}$LESIA, Observatoire de Paris, Universit{\'e} PSL, CNRS, Sorbonne Universit{\'e}, Universit{\'e} de Paris, 5 place Jules Janssen, 92195 Meudon, France\\
	$^{11}$ASTRON, Netherlands Institute for Radio Astronomy, Oude Hoogeveensedijk 4, 7991 PD, Dwingeloo, The Netherlands\\
	$^{12}$Department of Astrophysics/IMAPP, Radboud University Nijmegen, P.O. Box 9010, 6500 GL Nijmegen, The Netherlands\\
	$^{13}$Institute of Astrophysics, FORTH, N. Plastira 100, 70013, Heraklion, Greece\\
	$^{14}$Argelander Institut für Astronomie, Auf dem H{\"u}gel 71, 53117, Bonn, Germany\\
	$^{15}$Fakult{\"a}t f{\"u}r Physik, Universit{\"a}t Bielefeld, Postfach 100131, 33501 Bielefeld, Germany\\
	$^{16}$INAF - Osservatorio Astronomico di Cagliari, via della Scienza 5, 09047 Selargius (CA), Italy\\
	$^{17}$School of Physics, Faculty of Science, University of East Anglia, Norwich NR4 7TJ, UK\\
	$^{18}$Max-Planck-Institut f{\"u}r Gravitationsphysik (Albert Einstein Institut), Am M{\"u}hlenberg 1, 14476 Golm, Germany\\
	$^{19}$Jodrell Bank Centre for Astrophysics, Department of Physics and Astronomy, University of Manchester, Manchester M13 9PL, UK\\
	$^{20}$Laboratory of Gravitational Waves and Cosmology, Advanced Institute of Natural Sciences, Beijing Normal University at Zhuhai 519087, P.R.China\\
	$^{21}$Canadian Institute for Theoretical Astrophysics, University of Toronto, 60 St. George Street, Toronto, ON M5S 3H8, Canada\\
	$^{22}$Arecibo Observatory, HC3 Box 53995, Arecibo, PR 00612, USA\\
	$^{23}$Universit{\`a} di Cagliari, Dipartimento di Fisica, S.P. Monserrato-Sestu Km 0,700 - 09042 Monserrato (CA), Italy\\
	$^{24}$Laboratoire Univers et Th{\'e}ories LUTh, Observatoire de Paris, Universit{\'e} PSL, CNRS, Universit{\'e} de Paris, 92190 Meudon, France\\
	$^{25}$Institute for Gravitational Wave Astronomy and School of Physics and Astronomy, University of Birmingham, Edgbaston, Birmingham B15 2TT, UK\\
	$^{26}$Department of Astronomy, Peking University, Beijing 100871, P. R. China\\


\appendix

\section{Supplementary material}

\begin{table*}
\caption{Results of single-pulsar noise analysis for \eprise{} (EP) and \tn{} (TN) for DR1 in the same style as Table \ref{tab:spna_dr2_params}.}
\begin{center}
\def\arraystretch{1.5}
\begin{tabular*}{0.88\textwidth}{c||cc|cc||cc|cc}
\hline
 & $\log_{10} A_{\rm{RN}}$ & & $\gamma_{\rm{RN}}$ & & $\log_{10} A_{\rm{DM}}$ & & $\gamma_{\rm{DM}}$ & \\
Pulsar & EP & TN &  EP & TN &  EP & TN &  EP & TN \\
\hline \hline
J0613$-$0200    &  $-14.44_{-1.81}^{+1.3}$ & $-14.34_{-1.49}^{+1.13}$ &  $4.11_{-3.12}^{+2.69}$ &  $4.03_{-2.57}^{+2.64}$ & $-11.6_{-0.21}^{+0.14}$ &  $-11.6_{-0.15}^{+0.12}$ & $1.1_{-0.66}^{+0.87}$ & $1.11_{-0.53}^{+0.64}$ \\
\hline
J1012+5307    & $-12.99_{-0.2}^{+0.17}$ & $-12.99_{-0.16}^{+0.14}$ & $1.48_{-0.68}^{+0.86}$ &  $1.47_{-0.57}^{+0.64}$ & $-15.22_{-2.65}^{+3.24}$ & $-15.41_{-2.37}^{+3.17}$ & $2.8_{-2.67}^{+3.93}$ & $3.03_{-2.77}^{+3.48}$ \\
\hline
J1600$-$3053    & $-13.27_{-4.38}^{+0.2}$ & $-13.29_{-4.45}^{+0.21}$ & $1.38_{-1.05}^{+4.75}$ & $1.44_{-1.12}^{+4.91}$ & $-13.93_{-3.86}^{+2.48}$ & $-13.52_{-4.24}^{+2.08}$ & $2.0_{-1.76}^{+4.65}$
 & $1.92_{-1.65}^{+4.68}$ \\
\hline
J1713+0747    & $-15.01_{-1.91}^{+1.2}$ & $-15.03_{-0.92}^{+0.97}$ & $4.89_{-3.36}^{+1.97}$ & $5.09_{-2.05}^{+1.69}$ & $-11.69_{-0.11}^{+0.11}$ & $-11.7_{-0.08}^{+0.08}$ & $1.28_{-0.53}^{+0.56}$ & $1.22_{-0.4}^{+0.44}$ \\
\hline
J1744$-$1134    & $-14.21_{-3.38}^{+0.8}$ & $-14.0_{-3.0}^{+0.56}$ & $3.03_{-2.47}^{+3.61}$ & $3.01_{-1.78}^{+3.24}$ & $-11.79_{-0.22}^{+0.19}$ & $-11.83_{-0.2}^{+0.2}$ & $0.64_{-0.6}^{+0.88}$ & $0.58_{-0.54}^{+0.93}$ \\
\hline
J1909$-$3744    & $-13.9_{-3.61}^{+0.21}$ & $-13.9_{-3.62}^{+0.21}$ & $1.91_{-1.36}^{+3.67}$ & $1.91_{-1.35}^{+3.7}$ & $-14.87_{-2.98}^{+2.79}$ & $-14.94_{-2.91}^{+2.86}$ & $2.55_{-2.36}^{+4.14}$ & $2.58_{-2.39}^{+4.12}$ \\
\hline
\end{tabular*}
\label{tab:spna_dr1_params}
\end{center}
\end{table*}

\begin{table*}
\caption{$\log_{10} {\rm  BF}$ comparison between DR1 and DR2 in the same style as Table \ref{tab:bf_dr2}.}
\begin{center}
\def\arraystretch{1.5}
\begin{tabular*}{0.65\textwidth}{c|c||cc|cc}
\hline
\ & & DR1 DE438 & & DR2 DE438 & \\
Model ID & Model & \eprise{} & \ftwo{} & \eprise{} & \ftwo{} \\
\hline \hline
0 & PSR                 & $-$ & $-$ & $-$ & $-$ \\
\hline
1 & PSR + CURN          & $1.3$ & $1.2$ & $3.8$ & $3.6$  \\
\hline
2 & PSR + GWB           & $0.9$ & $0.8$ & $3.4$ & $3.2$  \\
\hline
3 & PSR + CLK           & $-0.3$ & $-0.3$ & $0.6$ & $0.8$  \\
\hline
4 & PSR + EPH           & $0.6$ & $0.3$ & $2.1$ & $2.2$  \\
\hline
5 & PSR + CURN + GWB    & $1.2$ & $0.5$ & $3.6$ & $3.7$  \\
\hline
6 & PSR + CURN + CLK    & $1.0$ & $0.4$ & $3.7$ & $3.4$  \\
\hline
7 & PSR + CURN + EPH    & $1.1$ & $0.5$ & $3.7$ & $3.4$  \\
\hline
\end{tabular*}
\label{tab:bf_dr1_dr2}
\end{center}
\end{table*}

\begin{figure*}
\centering
\includegraphics[width=0.42\textwidth]{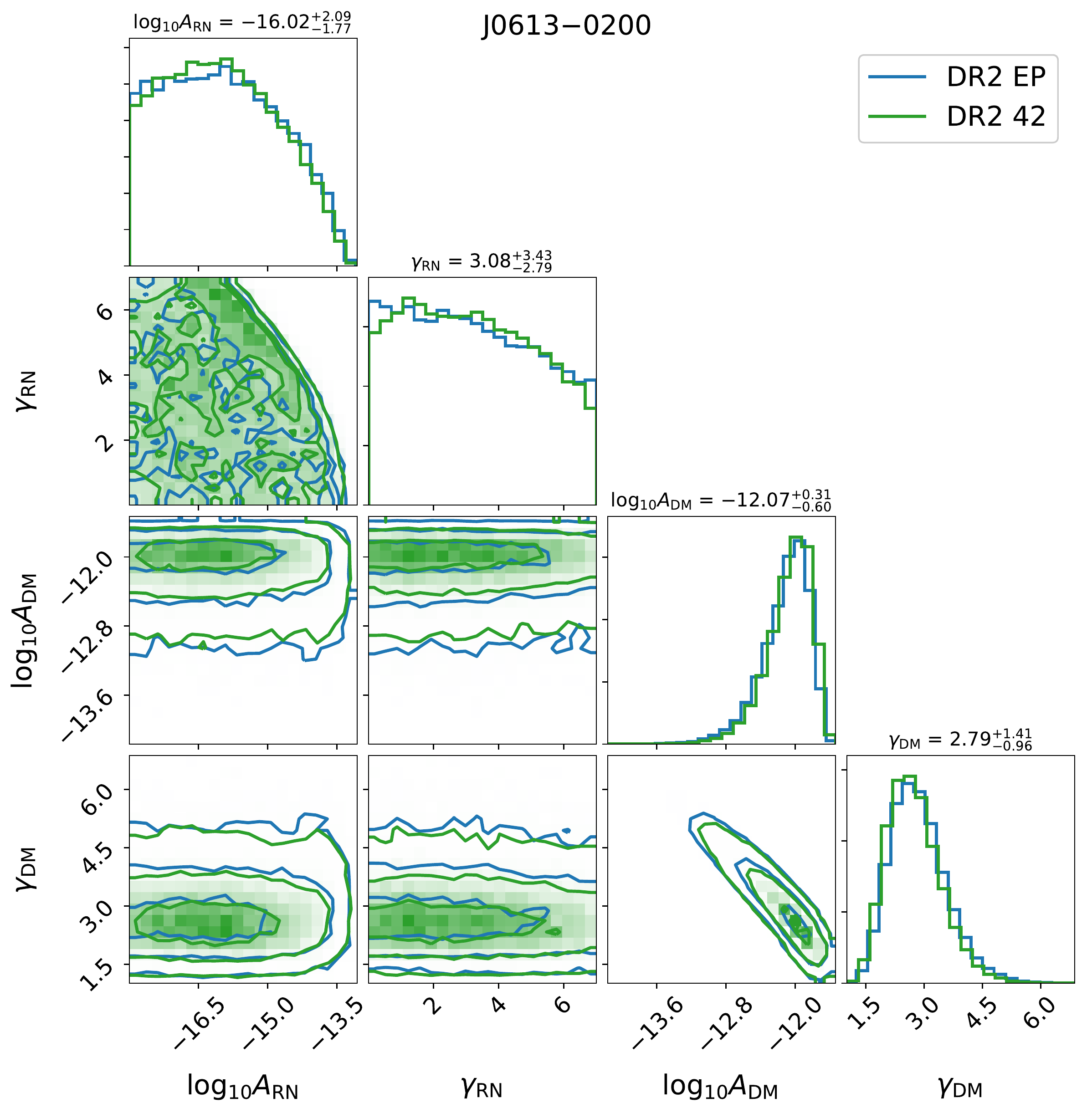} \hspace{0.6cm}
\includegraphics[width=0.42\textwidth]{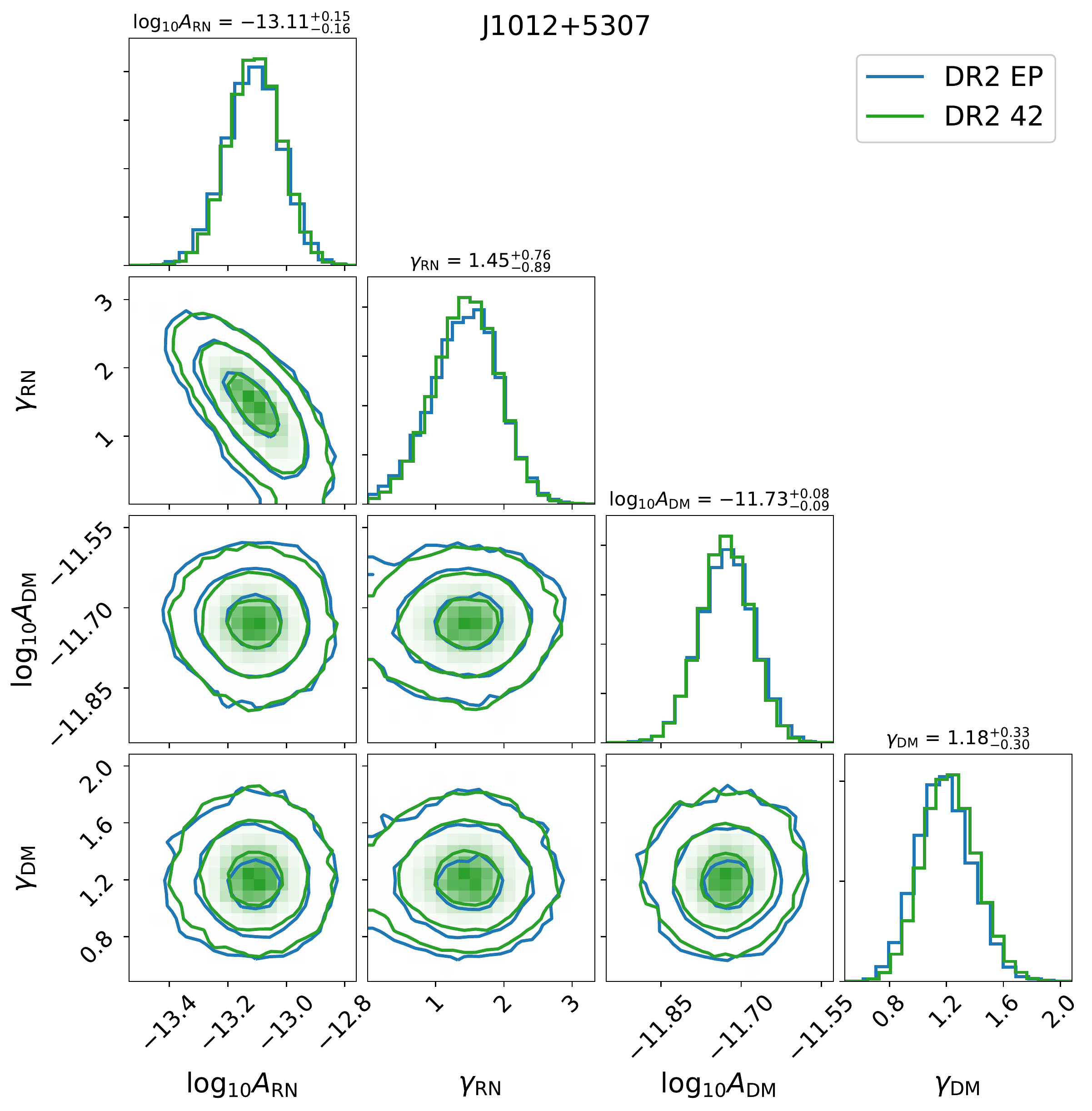} \\ \vspace{0.3cm}
\includegraphics[width=0.42\textwidth]{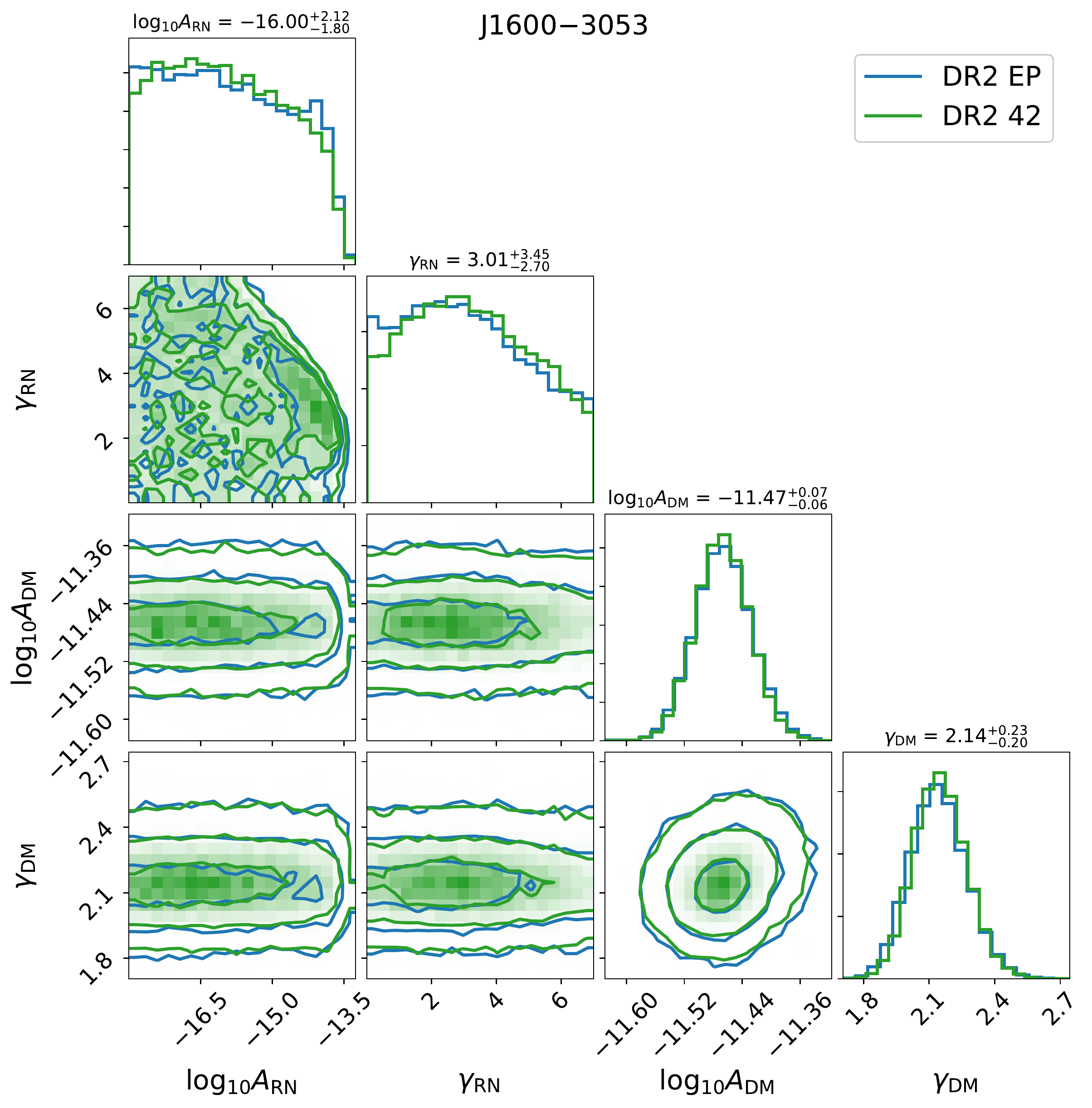} \hspace{0.6cm}
\includegraphics[width=0.42\textwidth]{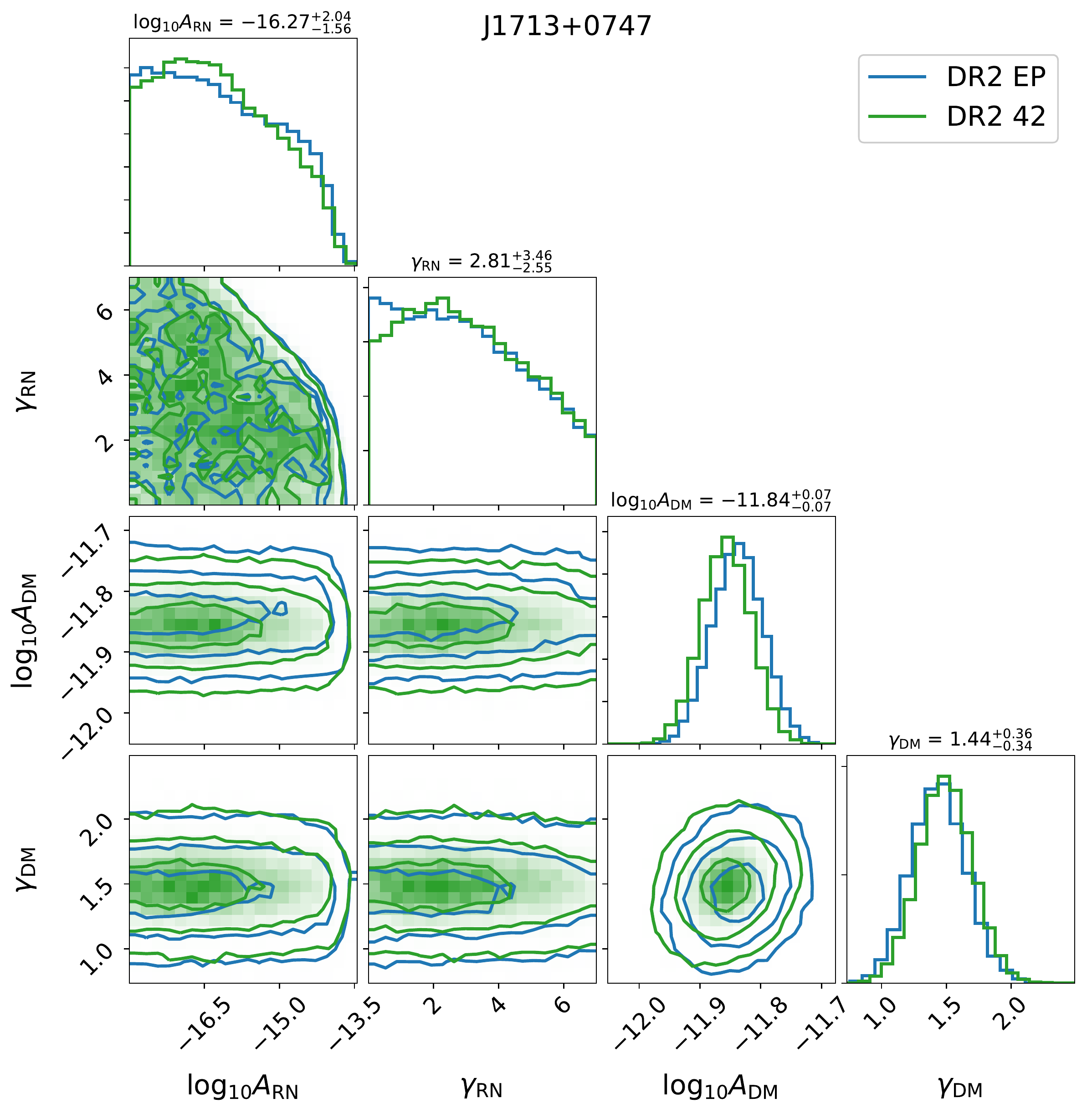} \\ \vspace{0.3cm}
\includegraphics[width=0.42\textwidth]{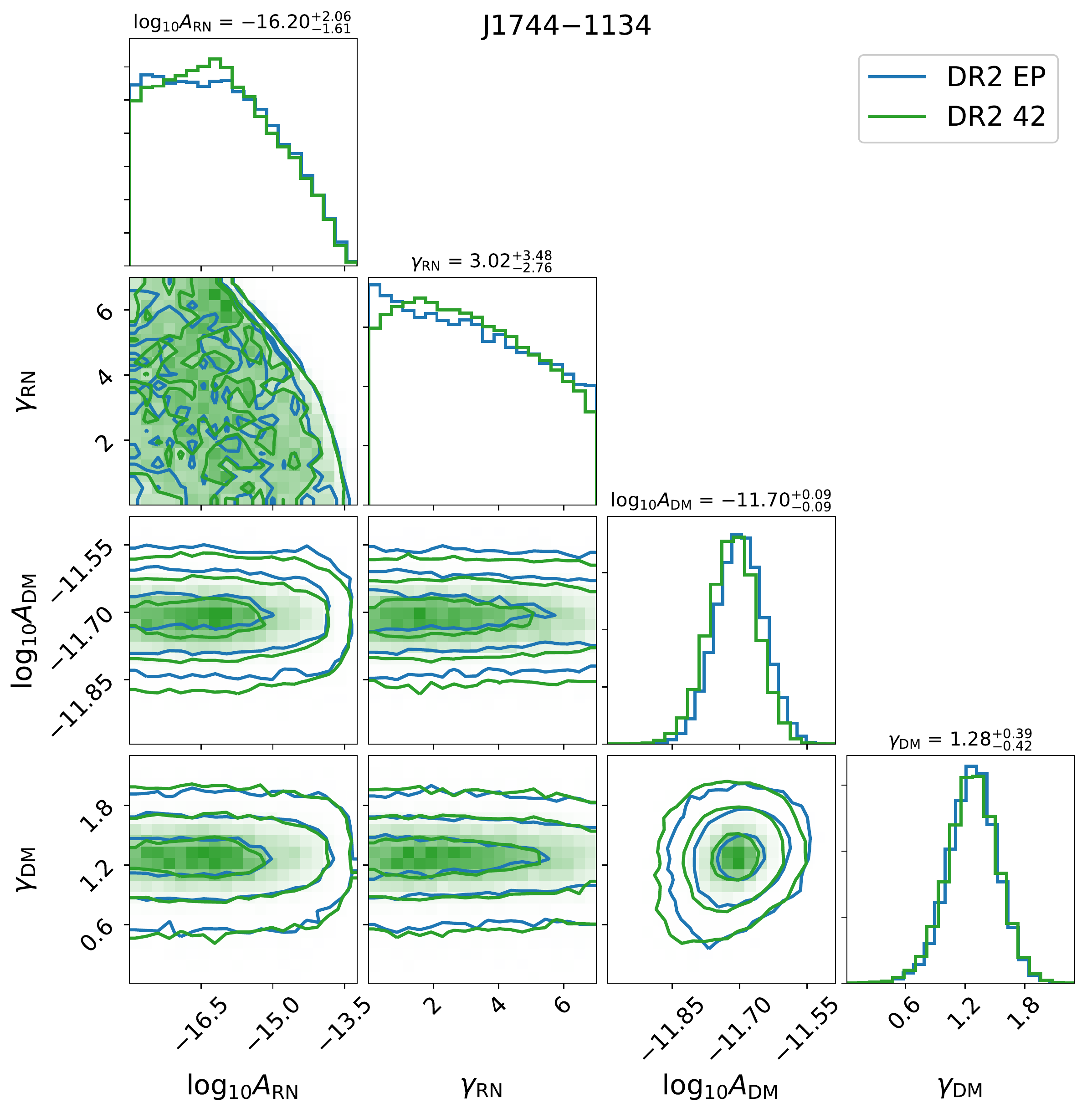} \hspace{0.6cm}
\includegraphics[width=0.42\textwidth]{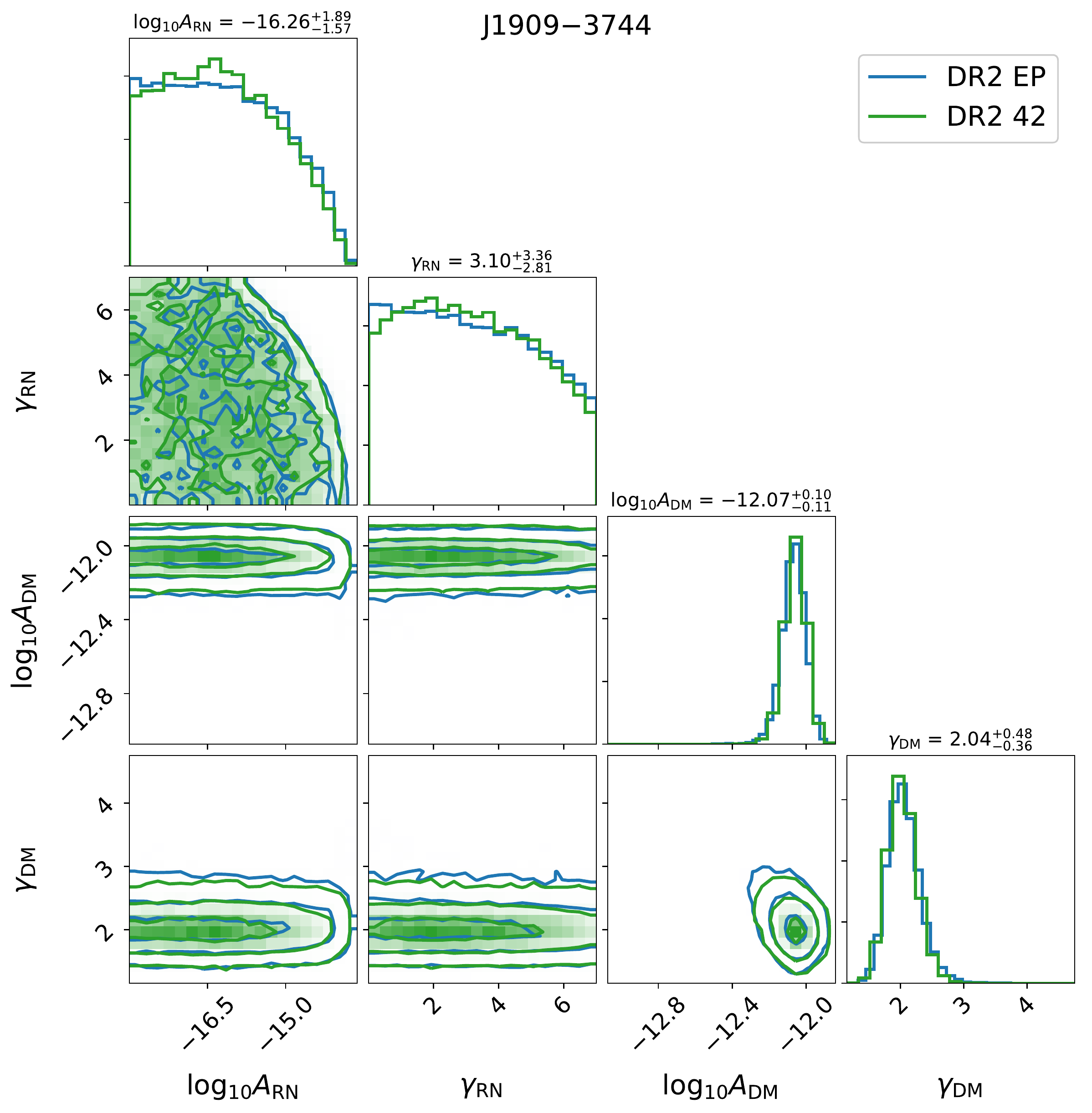}
\caption{Comparison of the parameter posterior distributions from the common uncorrelated red noise search between \eprise{} (EP) and \ftwo{} (42) split by pulsar noise parameters. The numbers indicate the 95\% credible regions from the EP analysis.}
\label{fig:cmp_psr}
\end{figure*}


\bsp	
\label{lastpage}
\end{document}